\documentclass[10pt,twocolumn,preprintnumbers,amsmath,amssymb,nofootinbib
,superscriptaddress]{revtex4}
\usepackage{graphicx,longtable,mathrsfs,color,array}
\usepackage[hidelinks]{hyperref}
\usepackage[usenames,dvipsnames]{xcolor} % colour
\usepackage{amssymb,amsmath,mathtools,mathrsfs,slashed} % maths
\usepackage{epsfig,subfigure,placeins,float} % plots
\usepackage{booktabs,longtable,ctable,multirow} % tables
\usepackage{exscale,relsize} % scale 
\usepackage[normalem]{ulem} % editing
\usepackage{enumerate}
\usepackage{times, mathptmx} % Times font
\usepackage{color}
\usepackage{hyperref}
\usepackage{graphicx}% Include figure files
\usepackage{color}
\usepackage{graphicx,graphics}

\newcommand{\be}{\begin{equation}}
\newcommand{\ee}{\end{equation}}
\newcommand{\bea}{\begin{eqnarray}}
\newcommand{\eea}{\end{eqnarray}}

\newcommand{\nn}{\nonumber}
  
\usepackage[stable]{footmisc}
\usepackage{color}

\def\dkmu2{\delta K_{\mu \nu}\delta K^{\mu \nu}}
\def\pmu2{  \phi_{\mu \nu}\phi^{\mu \nu}}

\begin{document}

\title{Inflation in a scale invariant universe.}
\author{Pedro G. Ferreira}
\email{pedro.ferreira@physics.ox.ac.uk}
\affiliation{Astrophysics, University of Oxford, DWB, Keble Road, Oxford OX1 3RH, UK}
\author{Christopher T. Hill}
\affiliation{Fermi National Accelerator Laboratory, P.O.Box 500, Batavia, Illinois 60510, USA}
\author{Johannes Noller}
\email{johannes.noller@eth-its.ethz.ch}
 \affiliation{Institute for Theoretical Studies, ETH Zurich, Clausiusstrasse 47, 8092 Zurich, Switzerland}
\author{Graham G. Ross}
\affiliation{Theoretical Physics, University of Oxford, 1 Keble Road, Oxford OX1 3NP, UK}
\date{Received \today; published -- 00, 0000}

\begin{abstract}
A scale-invariant universe can have a period of accelerated expansion at early times: inflation. We use a frame-invariant approach to calculate inflationary observables in a scale invariant theory of gravity involving two scalar fields - the spectral indices, the tensor to scalar ratio, the level of isocurvature modes and non-Gaussianity. We show that scale symmetry leads to an exact cancellation of isocurvature modes and that, in the scale-symmetry broken phase, this theory is well described by a single scalar field theory. We find the predictions of this theory strongly compatible with current observations.
\end{abstract}

%\pacs{98.80.-k, 98.80.Es, 95.36.+d, 95.36.+x}

\maketitle

\section{Introduction}
\label{sec:intro}
The evolution of our universe is simple to describe, yet difficult to explain. Current observations allow a picture in which there were two periods of accelerated expansion, one at very early times (dubbed inflation) and another at late times (i.e. today), separated by periods of radiation and matter domination. The hierarchy between the energy scales of the two regimes of accelerated expansion is extreme and difficult to understand in terms of our current knowledge of the interplay between particles, fields and gravity. Given this state of affairs, it is essential to find a consistent and simple explanation.

If one is to embrace inflation as an essential feature of the early universe (although one should, of course, countenance alternatives), it makes sense to explore alternative ideas which may explain the hierarchy of scales one encounters. A tried and tested approach is to invoke new symmetries which can naturally lead to such a hierarchy. In this paper we will explore one such symmetry - scale (or Weyl) invariance - which has been shown to lead to the type of behaviour we are seeking to understand \cite{Ferreira:2016vsc,Ferreira:2016wem,GarciaBellido:2011de,Rubio:2014wta,Bezrukov:2014ipa,Karananas:2016grc,Wetterich:1987fm,Kallosh:2013maa,Carrasco:2015rva,Quiros:2014wda,Kurkov:2016zpd,Karananas:2017zrg,Karam:2017zno,Rubio:2017gty,Karananas:2017mxm,Kannike:2016wuy,Einhorn:2017icw,Salvio:2017qkx,Salvio:2017xul,Salvio:2014soa}.

It has been shown that two scalar fields with a scale-invariant potential can be non-minimally coupled to gravity in such a way as to lead to a completely scale-invariant theory of the universe. While there are no dimensionful coupling constants, scale-symmetry is spontaneously broken and can generate a Planck mass, effective cosmological constant and particle masses. While, in the symmetry broken phase, dimensionful quantities emerge, the only meaningful, measurable quantities are ratios of dimensionful quantities which completely set by the dimensionless parameters of the underlying theory. A judicious choice of these parameters allows us to obtain two periods of accelerated expansion which are consistent with current observations. 

In this paper we will scrutinize the inflationary regime of the scale-invariant universe. Given that such a universe involves two scalar fields, one should expect a richer, more complex, phenomenology than a usual single field model. In particular, one should inspect the possible presence of isocurvature modes \cite{Gordon:2000hv} as well as non-negligible non-Gaussianity \cite{Bartolo:2004if}. The conventional approach for studying such models is to transform them from the Jordan frame into the Einstein frame to work out the properties of the scalar field evolution.  In this paper we will explore this phenomenology, using the frame-invariant approach of \cite{KP}. We will find that the mechanism of scale symmetry breaking greatly simplifies the calculations and that the final answer can be understood in terms of an effective single field model.

Our analysis extends previous related work in a number of ways: I) Using the analytic solutions for the scalar field evolution found in \cite{Ferreira:2016vsc}, we analyse various primordial observables in detail, finding good agreement with previous results, e.g. from \cite{Ferreira:2016vsc,Ferreira:2016wem,GarciaBellido:2011de,KP}, where overlap exists. We prove that, at next-to-leading order in slow-roll, isocurvature modes decouple completely in our scale-invariant setup, also away from the attractor solution (thus extending the related attractor solution result of \cite{GarciaBellido:2011de}). II) We explicitly derive the corresponding effective single field theory and show it leads to the same predictions. III) We extend previous results by computing predictions for the running of the tensorial spectral index and the non-Gaussian $f_{\rm NL}$ parameter(s).  IV) We work out predictions of the model in the recently developed frame covariant setup of \cite{KP}. We use this e.g. to investigate what the precise nature of the link between decoupling isocurvature mode(s) and scale invariance is and show that this is a consequence of working in a two-field scale invariant model.

This paper is structured as follows. In Section \ref{sec:model} we present the essential characteristics of the scale invariant universe with a particular emphasis on the inflationary regime; we recap the analytic solutions the field evolution, first found in \cite{Ferreira:2016vsc}. In Section \ref{sec:slowroll} we summarize the frame-invariant approach of \cite{KP}. In Section \ref{sec:iso} we explore the two field dynamics and the isocurvature sector to assess how close this theory is to single field dynamics. In Section \ref{sec:observables}  we calculate the observables - the various spectral indices, the amplitude of tensor modes and non-Gaussianity and show that we can also derive these results from an effective single scalar field theory. In Section \ref{sec:discussion} we discuss our findings.

%%%%%MODEL
\section{The Model}
\label{sec:model}
In this paper we will work with a model with two scalar fields,
$\phi^A \equiv (\phi_1,\phi_2)$,\footnote{Capital Latin letters are therefore field space indices that run from 1 to 2 (e.g. $A=1,2$). They are raised and lowered with a field-space metric, which we will introduce in the following section.} coupled to gravity. In the Jordan frame (in which we will present the results of this section) the action is given by: 
\begin{eqnarray}
S&=&\int d^4 x\sqrt{-g}[\tfrac{1}{2}M^2({\vec \phi}) R-\tfrac{1}{2}\sum_{A=1}^2\nabla_\mu\phi^A\nabla^\mu \phi^A - W({\vec \phi})]
\nn \\
\label{S}
\end{eqnarray}
where $M^2=-\frac{1}{6}\sum_{A=1}^2\alpha_A (\phi^A)^2$ and
$W({\vec \phi})   =  \sum_{A,B=1}^2\lambda_{AB}(\phi^A)^2(\phi^B)^2
$ and Einstein summation convention is {\it not} assumed.
This theory has no input mass scales and is conformally invariant if $\alpha_A=1$. 

The equations of motion are given by
\begin{eqnarray}
\sum_{B=1}^2\left[I_{AB}+\frac{\alpha_A\phi^A \alpha_B\phi^B}{6M^2}\right]\Box\phi_B={\cal X}_A
\end{eqnarray}
where
\begin{eqnarray}
{\cal X}_A=\frac{\alpha_A\phi^A}{6M^2}\sum_{B=1}^2(\alpha_B-1)({\dot \phi}^B)^2+\frac{4\alpha_A\phi^A}{6M^2}W+W_{,A}
\end{eqnarray}
and $A_{,X}=\partial A/\partial \phi^X$. 

This system has a conserved Noether current, $\nabla_\mu K^\mu=0$ where $K^\mu=\nabla^\mu K$ and
\begin{eqnarray}
K=\frac{1}{2}\sum_{A=1}^2(1-\alpha_A)(\phi^A)^2 \label{ellipse}
\end{eqnarray}
If we take $\phi^A$ to be functions of $t$ only and consider a homogeneous and isotropic metric of the form $g_{\alpha\beta}=(-1,a^2\delta_{ij})$, we have that
\begin{eqnarray}
{\ddot K}+3\left(\frac{\dot a}{a}\right){\dot K}=0
\end{eqnarray}
so that 
\begin{eqnarray}
K=c_1+c_2\int\frac{dt}{a^3(t)}
\end{eqnarray}

We see here one of the fundamental characteristics of this theory: scale invariance is spontaneously broken as $K$ settles down to a constant value, corresponding to an ellipses in the ${\vec \phi}$ plane. The value of $K$ is not set by the potential but by the initial value of ${\vec \phi}$ which makes this mechanism significantly different from the more conventional forms of spontaneous symmetry breaking - we have dubbed this particular mechanism {\it inertial symmetry breaking} \cite{Ferreira:2018itt}. Although ${\vec \phi}$ can still vary along the ellipse it is confined to that trajectory which is not invariant under scale transformations.

At late times, there is a fixed point on the ellipse, when ${\dot \phi}^A=0$ and 
\begin{eqnarray}
\frac{4\alpha_A\phi^A}{6M^2}W+W_{,A}=0
\end{eqnarray}
An explicit solution is
\begin{eqnarray}
\left(\frac{\phi_2}{\phi_1}\right)^2=\frac{\lambda_{11}\alpha_2-\lambda_{12}\alpha_{1}}{\lambda_{22}\alpha_1-\lambda_{21}\alpha_2}
\end{eqnarray}
We can see that the final, fixed-point, end state is set by the ratio of the coupling constants; any dimensionfull constants, such as the effective Planck mass, $M^2$, will depend on an arbitrary (or accidental) scale arising from the spontaneous breaking of scale symmetry.

A remarkable feature of this model is that the degree of freedom orthogonal to the constraint surface given by Equation \ref{ellipse} -- the dilaton -- completely decouples from the other degrees of freedom \cite{Ferreira:2016kxi}. To slightly belabour this point: given that the dilaton is the Goldstone boson of the broken symmetry, one might expect it to be derivatively coupled. In fact, it can be shown that  the scale-invariance of the theory ensures that the dilaton - the putative mediator of a fifth force - decouples from the matter sector, has only a kinetic term, and is thus unconstrained by laboratory or astrophysical effects \cite{Ferreira:2016kxi}. 
We will see that this fact will play a role when we study the evolution of perturbations in the inflationary regime.

Our focus, in this paper, will not be on the end-state but on a putative period of slow roll on the ellipse, before ${\vec \phi}$ settles down on the final fixed point. The equations of motion in this slow roll regime are given by 
\begin{eqnarray}
\sum_{B=1}^2\left[I_{AB}+\frac{\alpha_A\phi^A\alpha_B\phi^B}{6M^2}\right][-3H\phi^B]=\frac{4\alpha_A\phi^A}{6M^2}W+W_{,A}
\end{eqnarray}
If we assume that $W\simeq \lambda_{22}(\phi_2)^4$ we have
\begin{eqnarray}
\left( \begin{array}{c}
\frac{4\alpha_1\phi_1}{6M^2}W+W_{,1}\\
\frac{4\alpha_2\phi_2}{6M^2}W+W_{,2}\\
\end{array} \right)=
\frac{4\lambda_{22}\alpha_1\phi_1\phi_2^4}{6M^2}\left( \begin{array}{c}
1\\
-\frac{\phi_1}{\phi_2}\\
\end{array} \right)
\\ \nonumber
\end{eqnarray}

In this regime we can solve the equations of motion exactly \cite{Ferreira:2016vsc}. Defining $M_A^2=-\frac{\alpha_A}{6}(\phi^A)^2$ we have
\begin{eqnarray}
M_1^2&=& M^2_{E}e^{-\nu N_J}  \nonumber \\
M_2^2&=& M^2_{E}\left[1+\gamma(1-e^{-\nu N_J})\right] 
\label{phisols}
\end{eqnarray}
where $\nu=-\frac{4}{3}\alpha_1$, $\gamma=\frac{\alpha_2(1-\alpha_1)}{\alpha_1(1-\alpha_2)}$ and $N_J$ is the number of e-foldings until the end of inflation in Jordan frame.\footnote{In other words, we have implicitly defined $N_J = 0$ at the end of inflation.} We have shown that these analytical solutions are an exquisite approximation to the full equations of motion in the slow roll regime. We will work with this solution in all that follows in this paper (although we will at some point compare with numerical solutions).

We can obtain the dynamics of the Einstein frame scale factor $a_E$ and the corresponding Hubble rate $H_E$ through a conformal transformation of the form
\begin{eqnarray}
a_E &=& M(\phi_1,\phi_2)a \nonumber \\
H_E &=& H+\frac{\dot M}{M} \label{HE}
\end{eqnarray}
Thus we can reconstruct the Einstein frame quantities. The final piece in the dictionary is the transformation between Einstein and Jordan frame e-foldings which is given by
\begin{align} \label{NJNE}
N_E &= N_J + \ln\left(\frac{M_{f}}{M_{i}}\right), \nn \\
&= N_J + \frac{1}{2}\ln \left(\frac{2\alpha_1(1-\alpha_2)}{\alpha_1+\alpha_2-2\alpha_1 \alpha_2 + (\alpha_1 -\alpha_2)e^{-\nu N_J}} \right)
\end{align}
where $M_{f},M_{i}$ are the final and initial values of $M$ (at the end and start of inflation) respectively. One can implicitly solve to find $N_J(N_E)$ and in this way consistently map the solution \eqref{phisols} into the Einstein frame. Note that one can always uniquely relate the scalar field values at any given time to the corresponding $N_E$ and $N_J$. %While one can therefore obtain analogous expressions in physically equivalent frames in this way, we emphasise the particularly simple form the solution takes on in Jordan frame.  

{Scale invariant theories are particularly interesting because they provide a possible explanation for the hierarchical difference between the Planck scale and the electroweak scale, the scale invariance requiring vanishing masses until spontaneously broken. As originally constructed \cite{GarciaBellido:2011de,Rubio:2014wta,Bezrukov:2014ipa}  $\phi_2$ was taken to model the Higgs with an hierarchy of VEVs ${\phi_2\over\phi_1} \ll 1$. Thus $\phi_1$ is dominantly responsible for setting the Planck mass and $\phi_2$ sets the electroweak scale with the ``Higgs'' self-coupling $\lambda_{22}=O(1)$. In this limit one gets ``Higgs inflation'' with $|\alpha_2| \gg 1$ needed to have an acceptable scale of inflation\footnote{If $\phi_2$ does not model the Higgs it is possible for $\lambda_{22}$ to be small and in this case $\alpha_2$ need not be large \cite{Ferreira:2016vsc}.}.

The other scalar couplings $\lambda_{11}$ and $\lambda_{12}$ must be hierarchically small to allow for a small cosmological constant and to keep the Higgs mass at the electroweak scale. In the absence of gravity this ordering of couplings is natural due to the underlying shift symmetry of the  Weyl invariant scalar potential. This shift symmetry is broken by the ``Higgs'' coupling to the Ricci scalar and to determine whether the hierarchy survives requires a calculation of gravitational radiative corrections -- an issue in need of further elaboration.
}

It is possible to generalize the scale-invariant model to one with many scalar fields. The dynamics will be qualitatively similar: inertial symmetry breaking will occur but now the symmetry broken phase will lie on a (hyper-)ellipsoid and there will be richer dynamics to deal with. In Appendix \ref{sec:3D} we briefly touch on one such case to discuss a particular aspect related to the perturbations.
%%%%%Formalism
\section{Frame-invariant slow roll parameters}
\label{sec:slowroll}
There is a substantial literature on multi-field, inflationary perturbations in the slow roll regime \cite{Sasaki:1995aw,Gordon:2000hv,Elliston:2011et,Kaiser:2013sna,Achucarro:2012fd}. When the dynamics involves more than one field, the trajectory in field space will play a crucial role in how perturbations evolve and, in particular, whether the curvature perturbation is preserved on super-horizon scales or whether it varies, sourcing isocurvature perturbations. As shown in \cite{Gordon:2000hv} the curvature of the field trajectories plays a crucial role in the quantifying how isocurvature perturbations are sourced.

Over the past couple of decades, a more geometric approach has emerged in which the geometry of field space - through the field space metric that enters the definition of the kinetic term of the scalar field action - can be used to determine the evolution of perturbation in the case of multi-field inflation. In the case of non-minimal coupling, the favoured approach is to conformally transform to the Einstein frame and apply the standard slow roll formalism. A battery of ready available algorithms have been made available by a number of authors which numerically solve the transport equations and can be solved in the case of generic potentials and actions which lead to scalar field evolution which is sufficiently close to the slow-roll regime. In this paper we will follow a slightly different approach proposed in \cite{KP} (and foreshadowed by \cite{Gong:2011uw}) and consider a {\it frame invariant} formalism for calculating the inflationary observables (we have checked that we obtain the same results if we use the standard, Einstein-frame, approach and illustrate that in subsequent sections).

The fundamental quantities that one needs to consider are the frame-invariant metric\footnote{
Here we have assumed canonical kinetic interactions for the scalar fields of the form
$-\frac{1}{2}\delta_{AB}\nabla_\mu\phi^A\nabla^\mu \phi^B$. If the kinetic structure is non-trivial in field space, i.e. we have kinetic interactions of the form $-\frac{1}{2}k_{AB}(\vec \phi) \nabla_\mu\phi^A\nabla^\mu \phi^B$, then the frame-invariant metric becomes
\begin{eqnarray}
G_{AB}=\frac{k_{AB}}{M^2}+\frac{3}{2}\frac{M^2_{,A}M^2_{,B}}{M^4}
\label{kinS}
\end{eqnarray}.}
\begin{eqnarray}
G_{AB}=\frac{\delta_{AB}}{M^2}+\frac{3}{2}\frac{M^2_{,A}M^2_{,B}}{M^4}
\end{eqnarray}
and potential
\begin{eqnarray}
U=\frac{W}{M^4}.
\end{eqnarray}
Note that the frame-invariant metric is simply the field space-metric one obtains when transforming to the Einstein frame.

One can construct a covariant vector on field space, $X_A$, given by
\begin{eqnarray}
X_A=(\ln U)_{,A}
\end{eqnarray}
which is the frame field of the curvature perturbation or the tangent to the geodesics {\it in field space} traced out by the scalar field evolution. We can then construct the corresponding contra-variant vector by raising indices with $G^{AB}$. Furthermore, we can use $G_{AB}$ to construct the connection coefficients, $\Gamma^A_{\phantom{A}BC}$, which will go into the definition of a bona-fide covariant derivative; for example we have that
\begin{eqnarray}
\nabla_AX^B=X^B_{,\phi_A}+\Gamma^B_{\phantom{B}AC}X^C
\end{eqnarray}

We can then define the frame invariant potential slow roll parameters \cite{KP}
\begin{eqnarray}
{\bar \epsilon}_U&=&\frac{1}{2}X^A X_A  \label{slowrollparams1}
\end{eqnarray}
associated to the norm of the flow vector in field space and its directed derivatives along the flow:
\begin{eqnarray}
{\bar \eta}_U&=&-X^A(\ln {\bar \epsilon}_U)_{,\phi_A}  \nn \\
{\bar \xi}_U&=&-X^A(\ln {\bar \eta}_U)_{,\phi_A}
\label{slowrollparams2}
\end{eqnarray}
A defining feature of these slow-roll parameters as defined above, is that they reduce to the standard Hubble slow-roll parameters in the slow-roll approximation. 

In this regime, it is also important to define a set of
parameters that are crucial for evaluating the strength of the isocurvature perturbations. A leading parameter is the acceleration vector between paths in the geodesic flow
\begin{eqnarray}
\omega^A&=&X^B\nabla_B\left[\frac{X^A}{\sqrt{2{\bar \epsilon}_U}}\right] 
\label{isoparams1}
\end{eqnarray}
here given up to second order in the slow-roll parameters and which should be complemented by two additional parameters
\begin{eqnarray}
{\bar \eta}_{ss}&=&\frac{\omega^A\omega^B}{\omega^2}\left[ \nabla_AX_B+X_AX_B  \right] + {\tfrac{2}{3}\bar\epsilon_U R^A_{\phantom{A}A} }\nonumber \\
{\bar \eta}_{\sigma\sigma}&=&X^AX^B\left[ \nabla_AX_B+X_AX_B  \right], % \nonumber \\
\label{isoparams2}
\end{eqnarray}
where we have used the Ricci tensor $R_{AB}$ of our curved field-space. 
From these parameters (and especially from $\omega^A$) we can reconstruct how curved the trajectories are in field space and, in particular, what the transfer function that converts curvature perturbations into isocurvature perturbations is. 

In order to do so, we finally also need to promote the implicit definition of the number of e-foldings in \eqref{phisols} to a frame covariant one. Making use of the frame covariant time derivative ${\cal D}_t T \equiv \tfrac{d\phi^C}{dt}\nabla_C T$ (for any tensor $T$ -- see \cite{KP} for details), we can use Equation \ref{HE} to get
\be
{\cal H} \equiv {\cal D}_t a/a = H_E,
\ee
where $t$ is physical time and $H_E$ satisfies \eqref{HE}. Analogously we can then define a frame covariant e-folding number $dN = -{\cal H}dt$. Solving this equation, we have that the frame covariant e-folding number $N$ is given by
\be \label{Nsol}
N = N_E,
\ee
From Equation \ref{NJNE} this gives us $N$ as a function of $N_J$ or, inverting the relation, lets us express $N_J$ as a function of $N$ and as such yields an explicitly frame covariant version of \eqref{phisols}.

%%%%%ISO
\section{Isocurvature modes and the attractor}
\label{sec:iso}
Multi-field models generically produce entropy transfer between modes, leading to isocurvature effects on top of the standard adiabatic evolution  \cite{Gordon:2000hv}. This is particularly important on super-horizon scales, where the comoving curvature perturbation is conserved during adiabatic evolution, but evolves in the presence of isocurvature perturbations \cite{White:2012ya}. Before computing observables, it is therefore important to investigate whether isocurvature modes are present and impact the evolution of modes.

Isocurvature effects in two-field models of the type considered here can be parametrised by and encoded via the transfer functions $T_{\cal{R}\cal{S}}$ and $T_{\cal{S}\cal{S}}$  which are defined via \begin{eqnarray}
\left( \begin{array}{c}
{\cal R}\\
{\cal S}
\end{array} \right)=
\left( \begin{array}{cc}
1& T_{{\cal R}{\cal S}}\\
0 & T_{{\cal S}{\cal S}}  \end{array} \right)\left( \begin{array}{c}
{\cal R}_* \\
{\cal S}_* \end{array} \right)
\end{eqnarray}
where ${\cal R}$ and ${\cal S}$ are the curvature and entropy perturbations and $*$ denotes horizon exit at $N_*$. In multi-field models with $\geq 3$ fields, additional isocurvature modes are present and the above transfer functions get complemented by additional ones linking all neighbouring modes (i.e. each ${\cal S}_{(n)}$ and ${\cal S}_{(n+1)}$) - see \cite{KP} for details. Going back to the two-field context, a derived transfer angle $\Theta$ is defined by
\begin{eqnarray}
\cos \Theta=\frac{1}{\sqrt{1+T^2_{{\cal R}{\cal S}}}}.
\end{eqnarray} 
In integral form, the transfer functions can then be written
\begin{align}
 T_{\mathcal{RS}}(N_*,N)\ &=\  -\int_{N_*}^N dN'  \,  A(N') \, T_{\mathcal{SS}}(N_*,N')\; ,
 \nn \\
 T_{\mathcal{SS}}(N_*,N)\ &=\  \exp\left[ - \int_{N_*}^N  dN' \, B(N')\right]\; .
\label{transferInt}
\end{align}
Note that the e-folding number used here is the frame covariant one and $N_\star$ is defined to be positive. 
%so modes relevant for observables today are associated with $N_\star \sim 60$.  
For two-field models $A$ and $B$ satisfy \cite{Kaiser:2012ak}
\begin{align}
A &= 2\omega,
&B &= -2{\bar \epsilon}_U -{\bar \eta}_{ss}+{\bar \eta}_{\sigma\sigma}-\frac{4}{3}\omega^2,
\label{ABeqn}
\end{align}
where we have defined $\omega^2 = |\omega_A \omega^A|$, with indices raised and lowered with $G_{AB}$. In evaluating the isocurvature effects, let us first note that $\epsilon_U$ and $\bar \eta_{\sigma\sigma}$ and ${R^A}_A$ are well-defined, finite and generically non-zero expressions for our model. We have derived explicit expressions for these quantities, but will not require these here. The important quantity is $\omega$. 

To proceed we should first note that scale invariance imposes a set of consistency conditions on the quantities at play in the expressions of section \ref{sec:slowroll}. For example we have that
\begin{eqnarray}
\ln U(\lambda {\vec \phi})=\ln U({\vec \phi})
\end{eqnarray}
for arbitrary $\lambda$ which in turn leads to the constraint
\begin{eqnarray}
\frac{d\ln U(\lambda {\vec \phi})}{d\lambda}\left|_{\lambda=1}\right.=\phi^AX_A=0.
\end{eqnarray}
This immediately allows us to explicitly write out $X_A$ as
\begin{eqnarray}
\left( \begin{array}{c}
X_1\\
X_2
\end{array} \right)=\frac{X_1}{\phi_2}
\left( \begin{array}{c}
\phi_2\\
-\phi_1
\end{array} \right).
\end{eqnarray}
When indices are raised with $G^{AB}$ we get
\begin{eqnarray}
\left( \begin{array}{c}
X^1\\
X^2
\end{array} \right)\propto
\left( \begin{array}{c}
(\alpha_2-1)\phi_2\\
-(\alpha_1-1)\phi_1
\end{array} \right).
\end{eqnarray}
Interestingly, this is the orthogonal, contravariant vector to $\partial_AK$, i.e.  to the ellipse from equation \ref{ellipse}. We can define a unit vector ${\hat X}^A=X^2/\sqrt{2{\bar \epsilon}_U}$ and rescale $\omega^A$ such that
\begin{eqnarray} 
{\hat \omega}^A\equiv\frac{\omega^A}{\sqrt{2{\bar \epsilon}_U}}={\hat X}^B\nabla_B {\hat X}^A
\end{eqnarray}
There are a few properties to note about this expression. First of all, because of the structure in ${X}^A$ arising from scale invariance, there is no $\lambda_{AB}$ dependence in ${\hat X}_A$. Furthermore, we generally (and independently of scale invariance) have that ${\hat X}_A{\hat \omega}^A=0$
, which means that ${\hat \omega}^A\propto\phi^A$ for a scale invariant setup like ours. Putting everything together one can evaluate the proportionality constant and in fact explicitly show that  
\begin{eqnarray} 
{\hat \omega}^A=0,
\label{omega}
\end{eqnarray}
which means that ${\hat X}^A$ is a geodesic flow associated with the metric $G_{AB}$. 
This result can be seen as an extension of the result of \cite{GarciaBellido:2011de}, where an analogous turn-rate was shown to vanish for a subset of \eqref{S} on the attractor solution. Specifically, there it was shown that (on the attractor solution) the turn-rate vanishes for a potential $W = \tfrac{\lambda}{4}(\phi_2^2-\tfrac{\alpha}{\lambda}\phi_1^2)^2$, which (in the context of our \eqref{S}) is equivalent to assuming a specific choice of $\lambda_{11}$. We therefore emphasise that \eqref{omega} here holds for arbitrary $\lambda_{ij}$ and without assuming any specific solution (attractor or otherwise) -- it follows directly from \eqref{S} and \eqref{isoparams1}.  
The result \eqref{omega} can also be neatly interpreted in terms of the equation of motion for $\phi^A$, which can be written as
\begin{equation}
{\cal D}_t {\cal D}_t \phi^A + 3 {\cal H} ({\cal D}_t \phi^A) + f U^{,{\phi_A}} = 0,
\end{equation}
where we recall the definition ${\cal D}_t T \equiv \tfrac{d\phi^C}{dt}\nabla_C T$ for any tensor $T$. Since ${\hat X}^A \propto U^{,{A}}$, the statement that  
${\hat X}^A$ is a geodesic flow associated with the metric $G_{AB}$ becomes equivalent to the observation that the drag-term $- 3 {\cal H} {\cal D}_t \phi^A$ is aligned with the force term $f U^{,{A}}$.
We re-iterate that our starting expression for $\omega^A$ \eqref{isoparams1} was accurate up to second order in slow-roll, so the same is true for the above derivation. This is crucial, since both ${\hat \omega}^A=0$ and the alignment of drag-term and force term in the equations of motion are trivially true at first order in slow-roll parameters, but highly non-trivial at higher orders.

While we can understand the cancellation of ${\hat \omega}^A=0$ in a geometric way via the above reasoning, one may wonder whether this cancellation can also be related to another underlying feature. In fact, we have checked that one can add further dimensionless coefficients to the model, e.g. via $\lambda_{1112}\phi_1^3 \phi_2$ and/or $\lambda_{1222}\phi_1 \phi_2^3$ terms in the potential and the conclusion remains unchanged. This strongly suggests that dimensionless coefficients (by themselves) never contribute to $\omega^A$, i.e. that $\omega^A = 0$ is intimately tied to the scale-free nature of our model (at least for a two-dimensional field space -- see discussion below). Note that this changes as soon as any dimensionful coefficient is added. We have explicitly checked, that as soon as e.g. a constant (and of course dimensionful) Planck mass $M_{Pl}$ is added to the terms multiplying the Ricci scalar or a quadratic mass term is added (controlled by a new parameter $m^2$) or a sixth-order interaction such as $\phi_i^6/m^2$ is added, $\omega^A$ picks up non-zero contributions. Crucially all dimensionless parameters of the model then also enter the expression and affect $\omega^A$'s value, but in order to have a non-zero $\omega^A$ in the first place, the presence of at least one such dimensionful parameter is required.

What does this mean for the total isocurvature contributions for our model? From \eqref{transferInt} and \eqref{ABeqn}) we obtain that $T_{\cal{R}\cal{S}} = 0$ and $\Theta = 0$ as a direct consequence of $\omega=0$. In other words, no isocurvature effects affect observables related to the curvature mode (within the approximations we have used throughout, i.e. up to second-order in slow-roll).
Secondly note that $B$ is finite and generically non-zero\footnote{A calculation analogous to the above shows that $\bar \eta_{ss}$ is a finite, non-divergent quantity (the vanishing $\omega^2$ in the denominator is compensated for by factors in the numerator).}, meaning that an initially present isocurvature mode can still undergo a non-trivial evolution due to $T_{\cal{S}\cal{S}}$. However, this is of course decoupled from the curvature mode, given that $T_{\cal{R}\cal{S}} = 0$, and so the isocurvature mode can never be sourced by the curvature mode.

While the focus of this paper is the scale invariant model with two scalar fields, one has to consider the fact that this is a special case: the constraint is a one dimensional curve -- the ellipse \eqref{ellipse} -- on which the inflationary trajectory lies. Fluctuations {\it along} the ellipse correspond to adiabatic perturbations, fluctuations {\it orthogonal} to the ellipse correspond to isocurvature fluctuations. That orthogonal degree of freedom corresponds to the dilaton, which we have shown in \cite{Ferreira:2016kxi} completely decouples.
%
%But we also know that the degree of freedom which is orthogonal to the constraints corresponds to the dilaton which, as we have shown in \cite{Ferreira:2016kxi} completely decouples.
%
This means that we do not expect that particular isocurvature mode to be seeded or to interact with the adiabatic mode. Given that it is the only isocurvature mode in this theory, we recover what we found.

To confirm our intuition, we can generalize our analysis to the case of multi-scalar fields, where the situation is more complex. There the constraint surface is a hyper-ellipsoide in which the inflationary trajectory is embedded. Again, there will be an isocurvature mode associated to the dilaton, i.e. orthogonal to the surface, but now there will also be isocurvature modes lying on the constraint surface. These will {\it not} decouple from the adiabatic mode and can be seeded during inflation. The hallmark for this is that $\omega^A$ will not be zero in this case. As an example we have considered the case of three scalar fields with a set-up which is essentially equivalent to our model: $\alpha_1<\alpha_2,\alpha_3$ and the potential (which now consists of all quartic combinations of $\phi_1^2$, $\phi_2^2$ and $\phi_3^2$) is dominated by $\lambda_{22}\phi_2^4$. In Appendix \ref{sec:3D} we discuss this case in more detail, explicitly showing that $\omega^A$ and its norm are non-zero, which means isocurvature perturbations are clearly present in the case with more than $2$ fields.

%%%%%Obs
\section{Observables and Single Field Dynamics}
\label{sec:observables}

We are now ready to compute the observable predictions of our model. Let us quickly summarise the dynamical regime we are exploring. We are assuming that $W\simeq \lambda_{22}\phi_2^4$ during the inflationary regime. In  \cite{Ferreira:2016vsc} we showed that this was a well defined slow roll regime and allowed us to find the analytical solutions of Section \ref{sec:model}. Furthermore, we have that $|\alpha_1|\ll 1$ while $\alpha_2$ is unconstrained.

If we now turn to two-point functions of scalar and tensor perturbations, we are interested in the spectral index of scalar perturbations $n_S$, its running $\alpha_S$, the spectral index of tensor perturbations $n_T$, its running $\alpha_T$, and finally the tensor-to-scalar ratio $r$. Their frame-invariant definition is \cite{KP}
\begin{eqnarray}
n_S&=&1-2{\bar \epsilon}_U-{\bar \eta}_{U}-{\cal D}_N(1+T^2_{RS}) \nonumber \\
\alpha_S&=&-2{\bar \epsilon}_U{\bar \eta}_U-{\bar \eta}_U{\bar \xi}_U+
{\cal D}_N {\cal D}_N(1+T^2_{RS}) \nonumber \\
n_T&=&-2{\bar \epsilon}_U \nonumber \\
\alpha_T&=&-2{\bar \epsilon}_U{\bar \eta}_U \nonumber \\
r&=&16{\bar \epsilon}_U\cos^2\Theta,
\label{genobs}
\end{eqnarray}
where ${\cal D}_N$ is the frame covariant derivative wrt. $N$, but since we have already seen that the transfer function $T_{RS}$ vanishes in our setup, all terms involving ${\cal D}_N$ drop out trivially and $\cos^2 \Theta = 1$. Note that we therefore trivially obtain the consistency relation $r = -8 n_T$.

Making use of \eqref{phisols}, we accordingly obtain exact expressions for all these observables. Expanding up to leading-order in $\alpha_1$ for each parameter, we find:\footnote{Note that, in this small $\alpha_1$ expansion, we have not expanded the exponential $e^{-\nu N_J}$, since it can be order one even if $|\alpha_1| \ll 1$.}
\begin{align}
n_S &= 1 + \frac{4\alpha_1 (e^{-\nu N_J}+1)}{3(1 - e^{-\nu N_J})} +  {\cal O}(\alpha_1^2), \nn \\
r &= \frac{64 \alpha_1^2 (\alpha_2 - 1) e^{-\nu N_J}}{3 \alpha_2 (e^{-\nu N_J} - 1)^2} + {\cal O}(\alpha_1^3), \nn \\
\alpha_S &=- \frac{32 \alpha_1^2 e^{-\nu N_J}}{9 (e^{-\nu N_J}-1)^2} + {\cal O}(\alpha_1^3), \nn \\
n_T &= -  \frac{8 \alpha_1^2 (\alpha_2 - 1) e^{-\nu N_J}}{3 \alpha_2 (e^{-\nu N_J} - 1)^2}  + {\cal O}(\alpha_1^3) , \nn \\
\alpha_T &= - \frac{32 \alpha_1^3 (\alpha_2 - 1) e^{-\nu N_J} (1 + e^{-\nu N_J})}{9 \alpha_2 (e^{-\nu N_J} - 1)^3}  + {\cal O}(\alpha_1^3), & &
\label{obsExp}
\end{align}
We have checked that, with the fiducial parameter values of \cite{Ferreira:2016wem}, these expressions are accurate at roughly percent level (when compared with the full expressions). Note that $N_J$ here (in the spirit of frame covariance) should be seen as a function of $N$. This can be obtained by inverting \eqref{Nsol}, which at leading-order in $\alpha_1$ becomes
\be \label{Nsol2}
N = N_J + \frac{1}{2} \ln\left(\frac{2\alpha_1 (\alpha_2-1)}{\alpha_2(e^{-\nu N_J} - 1 )}\right)
%+ \frac{\alpha_1 (1 - 2 \alpha_2 + e^{-\nu N_J})}{2 \alpha_2 (e^{-\nu N_J}-1)} 
+ {\cal O}(\alpha_1).
\ee
Also, here and in what follows, we are focusing on the modes relevant for observables today, by picking a fiducial $N_J \sim 60$. Using \eqref{Nsol2}, one can show this corresponds to $N \sim 58.5$.\footnote{Incidentally this is precisely in the parameter range explored by \cite{GarciaBellido:2011de}, which corresponds to $57 \lesssim N \lesssim 59$.}

These results extend, and are also completely consistent with, those found in  \cite{Ferreira:2016vsc}, where calculations were done using the $H(N)$ formalism, in the Einstein frame (see also \cite{GarciaBellido:2011de}). It is instructive to pursue this further. As we saw in Section \ref{sec:iso}, isocurvature perturbations are zero upto, at least, $2^{\rm nd}$ order which means that the there are no perturbations orthogonal to the field trajectory. One might have guessed that would be the case, given that the field is evolving along the scale symmetry broken locus of field space, i.e. the ellipse of Equation \ref{ellipse} but this doesn't immediately follow; the trajectory along the ellipse has curvature which one might naively associate with normal forces and thus isocurvature perturbations. Given that this is not the case (due to the way in which the $\phi^A$ map onto curvature and isocurvature modes) and $\omega^A=0$, we can simplify the analysis considerably by reducing the theory to a single field model.

Substituting the solutions for $\phi^A$ \eqref{phisols} into the ellipse equation \eqref{ellipse} and (without loss of generality) setting $M_E$ to unity in what follows, we find that
\begin{equation}
K = 6 -  \frac{3}{\alpha_1} -  \frac{3}{\alpha_2}.
\end{equation}
Solving for the ellipse, we can therefore express the whole theory in terms of a single degree of freedom, which we choose to be $\phi \equiv \phi_2$:
\begin{eqnarray}
S&=&\int d^4 x\sqrt{-g}[\frac{1}{2}\hat M^2({\phi}) R-\frac{\hat k (\phi)}{2}\nabla_\mu\phi\nabla^\mu \phi - \hat W({\phi})],
\nn \\
\label{Ssingle}
\end{eqnarray}
where we see explicitly that the single field formulation comes at the expense of introducing a non-canonical kinetic term. The model functions are given by
\begin{align}
\hat M^2 &= \frac{{\alpha_{2}} {\phi}^2 + {\alpha_{1}} (2 K -  {\phi}^2)}{6 ({\alpha_{1}}-1)}, \nn \\
\hat k &= - \frac{2 K(1-  {\alpha_{1}}) + ( {\alpha_{2}} - 1) ({\alpha_{2}} - {\alpha_{1}}) {\phi}^2}{({\alpha_{1}} - 1) \bigl(2 K + ({\alpha_{2}} - 1) {\phi}^2\bigr)} \nn \\
\hat U &= \frac{\hat W}{\hat M^4} = \frac{36 ( {\alpha_{1}}-1)^2 {\lambda_{22}} {\phi}^4}{\bigl({\alpha_{2}} {\phi}^2 + {\alpha_{1}} (2 K -  {\phi}^2)\bigr)^2}.
\label{modfunc}
\end{align}
Recalling the definition of the frame-invariant metric in the presence of non-trivial kinetic terms for the scalar(s) \eqref{kinS} and noting that the field space metric is a simple scalar function in the case of a one-dimensional field space as we are considering here, we have 
\begin{equation}
\hat G = \frac{12 ({\alpha_{1}}-1) K \bigl(2{\alpha_{1}} K + ({\alpha_{1}} -  {\alpha_{2}}) ( {\alpha_{2}}-1) {\phi}^2\bigr)}{\bigl(2K + ({\alpha_{2}}-1) {\phi}^2\bigr) \bigl(2{\alpha_{1}} K + ({\alpha_{2}} - {\alpha_{1}}) {\phi}^2\bigr)^2}.
\end{equation}
Expressed in this way, we have $\hat G = G_{AB}$ and consequently $G^{AB} = \hat G^{-1}$.
and can express the first two slow-roll parameters as 
\begin{align}
{\hat \epsilon}_{\hat U} &= \frac{\hat U_{,\phi}^2}{2 G {\hat U}^2} , 
&{\hat \eta}_{\hat U} &= -\frac{{\hat \epsilon}_{\hat U,\phi} \hat U_{,\phi}}{{\bar \epsilon}_{\hat U} G \hat U}
\end{align}
Evaluating this and expanding in $\alpha_1$, we obtain precisely the same expressions for $n_S$ and $r$ as in \eqref{obsExp}. In fact, we have explicitly checked that the two approaches yield identical predictions up to eighth order in $\alpha_1$.

\begin{figure}[htbp]
\begin{center}
\includegraphics[width=\linewidth]{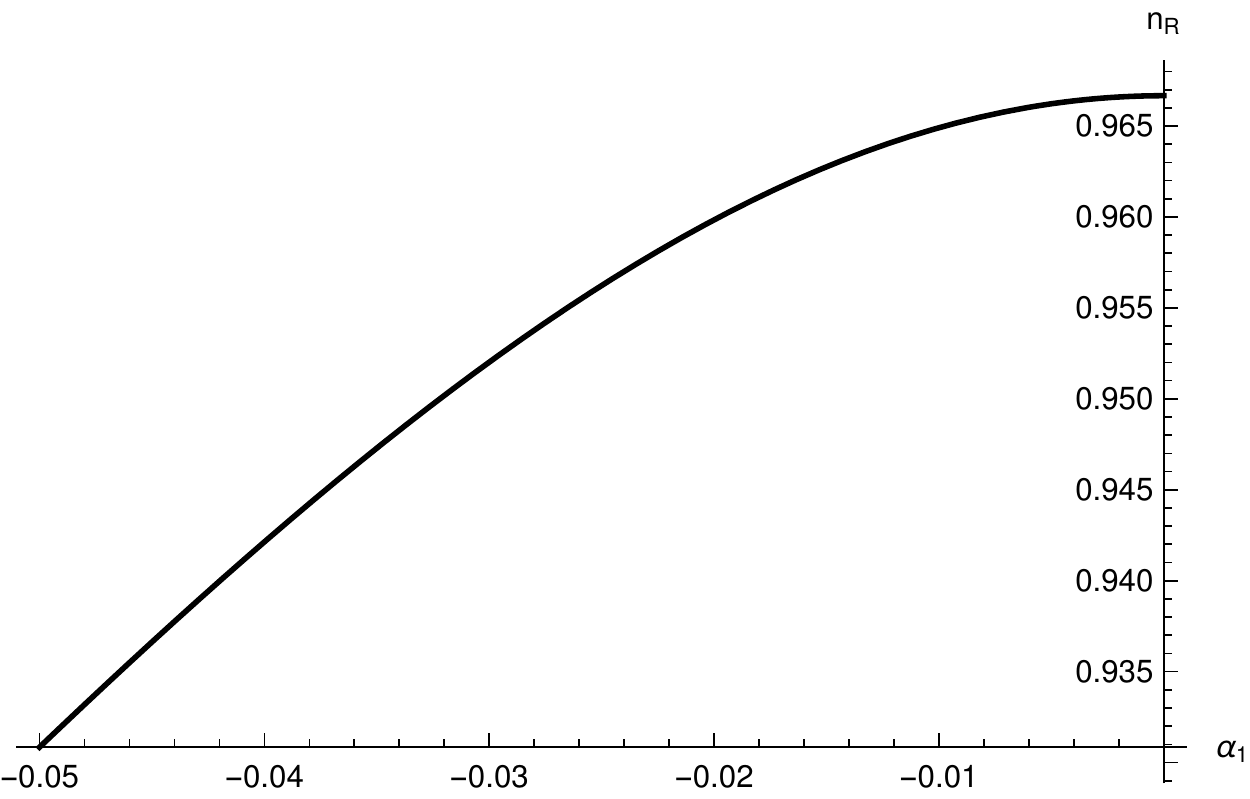}
\end{center}
\caption{Plot of $n_S$ vs. $\alpha_1$. Note that the spectral index of scalar perturbations does not depend on $\alpha_2$ at leading order in $\alpha_1$, which is unlike the result for $r$ above. \label{fig1}}
\end{figure}
\begin{figure}[htbp]
\begin{center}
\includegraphics[width=\linewidth]{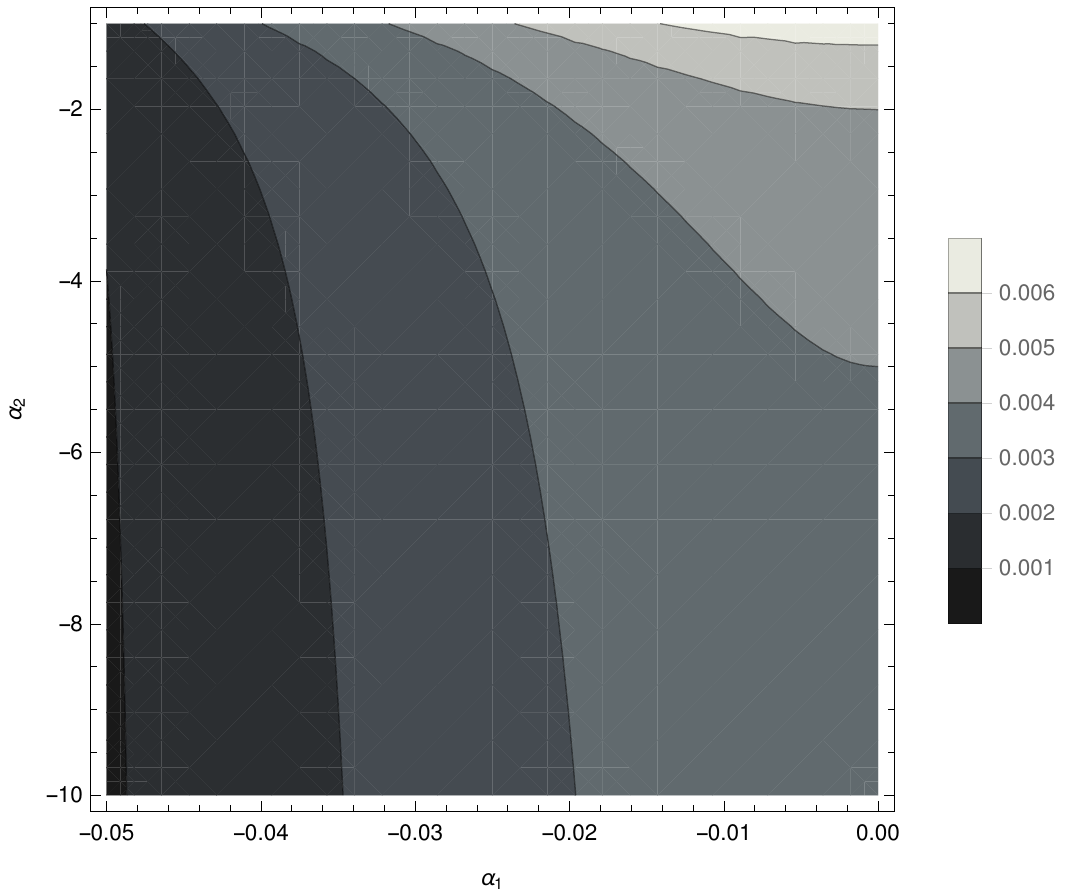}
\end{center}
\caption{Contour plot of $r$ vs. $\alpha_2$ and $\alpha_1$. Note that, as far as $r$ is concerned and at least for the values shown here, all parameter values give observationally consistent predictions. \label{fig2}}
\end{figure}

We can now focus on the actual values of the observables. The main observables, i.e. the ones for which we have the tightest constraints are $n_S$ and $r$. In Figure \ref{fig1} we can see that, for sufficiently small values of $\alpha_1$, $n_S\simeq 0.96$, i.e. it lies comfortably within the observational constraints from the Planck data \cite{PlanckCosmo2015}. In fact, given that $n_S$ is solely dependent on $\alpha_1$ we can immediately convert current constraints on $n_S$ (e.g. $n_S=0.9652\pm0.0047$) into  constraints on 
$\alpha_1$:
\begin{eqnarray}
 |\alpha_1| < 0.019 \label{alpha1Con}
\end{eqnarray}
Note that there is an upper bound  on $n_S$ for $\alpha_1\rightarrow 0$ such that $n_S<1-2/N\simeq 0.97$.

In Figure \ref{fig2} we can see that we naturally obtain a small value of $r$, well within current constraints. A conservative expression comes from taking $\alpha_1\rightarrow 0$:
\begin{eqnarray}
r\simeq \frac{12}{N^2}\frac{(\alpha_2-1)}{\alpha_2}\simeq \frac{1}{300}\frac{(\alpha_2-1)}{\alpha_2}.
\end{eqnarray} 
Current constraints on $r<0.07$ lead to a conservative bound on $\alpha_2$ such that
\begin{eqnarray}
\alpha_2< -0.048.
\end{eqnarray}
Given the constraint on $\alpha_1$ \eqref{alpha1Con}, there is interestingly also a lowest value for $r$ in our model, namely $r > 0.0026$.
The other observables are, currently, unconstrained but could in principle be measured with future CMB measurements (in the case of $n_T$ and $\alpha_T$) and high redshift 21 cm missions that can probe small wavelengths in the linear regime (in the case of $\alpha_S$). 
The numerical predictions our model makes for these parameters are the following:
$ \alpha_S \sim - 5 \cdot 10^{-4} \pm 10 \%$, depending on the precise value of $\alpha_1$. Note that, at leading order in $\alpha_1$, $\alpha_S$ is independent of $\alpha_2$, just like $n_S$.  $n_T$ and $\alpha_T$ both do have explicit dependence on $\alpha_2$ and $\alpha_1$. However, within the allowed range for $\alpha_1$ \eqref{alpha1Con}, the value for $n_T$ is well recovered by the limiting expression as $\alpha_1 \to 0$ and we then find
\begin{equation}
n_T \sim - 4 \cdot 10^{-4} \cdot \frac{(\alpha_2 - 1)}{\alpha_2}.
\end{equation}
For $\alpha_T$, considering the limiting expression as $\alpha_1 \to 0$ gives even more accurate results (due to the extra $\alpha_1$ suppression factor -- see equation \eqref{obsExp}) and we there obtain
\eqref{obsExp}
\begin{equation}
\alpha_T \sim - 1.4 \cdot 10^{-5} \cdot \frac{(\alpha_2 - 1)}{\alpha_2}.
\end{equation}

Finally we can also investigate signatures of the scale-invariant model beyond the 2-point function. While an exploration of the full bi- and trispectrum is beyond our scope, the local non-Gaussian parameter $f_{NL}$ provides an observable of particular interest, since it is strongly suppressed in single field models \cite{Maldacena:2002vr,Creminelli:2004yq} and can therefore provide a smoking-gun for multi-field dynamics, if sizeable enough to be measured.
Focusing on this local limit, one can then obtain the following expression \cite{Elliston:2012ab,Byrnes:2012sc} 
\begin{align}
f_{NL}^{\rm local} &\approx \frac{5}{6}\frac{N^{,A}N^{,B}(\nabla_A\nabla_BN)}{(N_{,C}N^{,C})^2},
\label{fnlN}
\end{align}
where $N$ is the frame covariant number of e-folds, as before. Note that this expression is essentially a (covariantised) version of the standard $\delta N$ expression for $f_{NL}$ \cite{Lyth:2005fi}, which there corresponds to a quasi-local configuration for the bispectrum (close, but not identical, to the local one -- cf. the discussion in \cite{Kenton:2015lxa}).
Taking \eqref{fnlN} and noting that one can write $N_{,A} = U U_{,A}/(U_{,B}U^{,B})$ \cite{KP}, after some algebra we then find that $f_{NL}^{\rm local}$ can in fact succinctly be expressed as
\begin{equation}
f_{NL}^{\rm local} \approx - \frac{5}{6}\frac{X^{A} X^{B} \nabla_{B}X_{A}}{(X_{C} X^{C})}= \frac{5}{12} \bar \eta_U.
\label{fnlX}
\end{equation}
Taking the same approach as for the other observables considered above, we can expand in $\alpha_1$ and find the following highly accurate expression
\begin{equation}
f_{NL}^{\rm local} \approx \frac{5 \alpha_1 (e^{-\nu N_J}+1)}{9 (e^{-\nu N_J}-1)} + {\cal O}(\alpha_1^2)\simeq \frac{5}{12}(1-n_S)\simeq \ {\rm few} \times 10^{-2}
\label{fnlns}
\end{equation}
where agreement between this expression to leading order in $\alpha_1$ and the full expression \eqref{fnlX} holds down to sub one-percent level. Phrasing it in terms of $n_s$ reproduces the (single-field) relation of \cite{Acquaviva:2002ud}, which is of course expected, given the existence of our effective single field description \eqref{Ssingle}. The $(n_S - 1)$ suppression in \eqref{fnlns} then also follows from the well-known consistency relations for the 3-point function \cite{Maldacena:2002vr,Creminelli:2004yq}. Finally note that we find $c_s = 1$ in the effective single field picture \eqref{Ssingle}, due to the independence of the model functions \eqref{modfunc} on derivatives of $\phi$. This immediately allows us to conclude that no sizeable equilateral non-Gaussianity is present in our model either, since for general single-field models $f_{NL}^{\rm equil} \lesssim 1/c_s^2$.

%%%%%Discussion
\section{Discussion}
\label{sec:discussion}
In this paper we have calculated the inflationary observables for inflation in a scale-invariant universe. While previous calculations had been undertaken in the Einstein frame under the assumption of single field evolution, we chose to consider the full multifield model in a scale invariant formalism. This allowed us to prove that, up to second order in slow roll, no isocurvature perturbations were generated in the inflationary regime. We showed that this was a particular feature of the two field model we are considering in this paper and can be understood quite simply: the isocurvature mode is orthogonal to the constraint ellipse and thus we can identify it with the dilaton. As we have shown before, the dilaton completely decouples from the other degrees of freedom. Our result reinforces the fact that the (effective) single field approach is an excellent approximation.

Nevertheless, we persisted with the calculation taking into account both fields and found a set of analytic expressions for the inflationary observables: $r$, $n_S$, $\alpha_S$, $n_T$ and $\alpha_T$. These expression are accurate at the sub-percent level; $r$ and $n_S$ are in exact agreement with those found in \cite{Ferreira:2016vsc}. As a final cross-check, we explicitly reduced the system to the dynamics of a single field by solving for the constraint in Equation \ref{ellipse}. Again, we recovered the same analytic results as we had determined in the multifield case, reinforcing the fact that isocurvature perturbations are completely absent. Finally, we assessed the level of non-Gaussianity in this model and found it to be small, of order $f^{\rm local}_{NL}\sim10^{-2}$ and well within the current observationally allowed range.

Our calculations have confirmed that inflation in a scale-invariant universe is a completely viable model for the origin of structure, leading to acceptable observables. Furthermore it is fundamentally well motivated; in future attempts at cosmological constraints one is in a position to consider priors on the fundamental parameters as opposed to on the observables (such as $r$ and $n_S$). Our results also reinforce the point made in \cite{Kaiser:2013sna}: if we are to accept inflation as the theory that explains the seeds for structure, then current data is strongly pushing us to have to accept non-minimal couplings. This is a striking statement about the fundamental structure of gravity and a further incentive to consider theories such as the one discussed in this paper.

In this paper we have not touched on other fundamental issues in inflation model building that need to be addressed: how did the inflationary regime begin, how fine-tuned are the initial conditions? In the scale-invariant model, these questions are intimately tied to the inertial symmetry breaking that occurs and leads the fields to lie on the constraint surface (the ``ellipse''). The slow roll conditions are naturally enforced on a large region of the ellipse but whether, for a general set of initial conditions, the fields naturally end up in that region, remains to be seen.

\textit{Acknowledgements ---} We are extremely grateful for discussions with Ana Achucarro, David Kaiser, Sotirios Karamitsos, David Mulryne and David Wands. PGF acknowledges support from 
Leverhulme, STFC, BIPAC and the ERC. JN acknowledges support from Dr.\ Max R\"ossler, the Walter Haefner Foundation and the ETH Zurich Foundation. Part of this work was done at Fermilab, 
operated by Fermi Research Alliance, LLC under Contract No. DE-AC02-07CH11359 with the United States Department of Energy.

%===============================================================================
\appendix
\section{Geometry of field space}
\label{sec:geometry}
For ease of notation we define $f \equiv M^2= -\tfrac{1}{6} \sum_A \alpha_A (\phi^A)^2$. We will need are the following
\begin{eqnarray}
f &=& -\tfrac{1}{6} \sum_A \alpha_A (\phi^A)^2 \nonumber \\
f_{,B} &=& -\tfrac{1}{3} \sum_A \alpha_A \phi^A \delta_B^A = -\tfrac{1}{3} \alpha_B \phi^B \nonumber \\
f_{,B,C} &=&  -\tfrac{1}{3} \alpha_B \delta^B_C.
\end{eqnarray}
It is useful to define
\begin{equation}
F \equiv \sum_D f_{,D}f_{,D} = \tfrac{1}{9} \sum_D \alpha_D^2 (\phi^D)^2.
\end{equation}
Inserting the above expressions into the definition of the field-space metric, we find
\begin{align}
G_{AB} = \frac{1}{f}\left(\delta_{AB} + \frac{1}{6 f} \alpha_A \alpha_B \phi^A \phi^B \right),
\end{align}
and for the inverse field-space metric we ihave
\begin{align}
G^{AB} = f \delta^{AB} -  \frac{3 f^{,A} f^{,B}}{2 \left(1 + \frac{3 F}{2 f}\right)} =  f \delta^{AB} -  \frac{\alpha_A \alpha_B \phi^A \phi^B}{6 \left(1 + \frac{3 F}{2 f}\right)}.
\end{align}
We can now also express the connection $\Gamma$ as
%\begin{widetext}
\begin{align}
\Gamma^A_{\phantom{A}BC} &= \frac{2 f \delta_{BC} f^{,A} 
- (3 F + 2 f) ( \delta^{A}_{C} f_{,B}+\delta^{A}_{B} f_{,C})  + 6 f f^{,A} f_{,C,B}}{2 f (3 F + 2 f)} \nonumber \\
&= \frac{(F + \tfrac{2}{3} f) (\delta^{A}_{C}\alpha_B \phi^B +\delta^{A}_{B}\alpha_C \phi^C) + \tfrac{2}{3} f \alpha_A \phi^A (\alpha_C-1) \delta_{BC}}{2 f (3 F + 2 f)}.
\end{align}
%\end{widetext}
Finally we have that
\begin{align}
X_C &= \frac{U_{,C}}{U}= \frac{2}{U f^2}\sum_{A,B}\lambda_{AB} \phi^A  (\phi^B)^2 \delta^A_C + \frac{2}{3}\frac{\alpha_C\phi_C}{f}.
\end{align}

%\widetext

%%APPENDIX 3D
\section{Isocurvature modes for a scale-invariant 3-field theory}
\label{sec:3D}

Here we briefly discuss a scale invariant 3-field model analogous to the 2-field model presented in the main body of the paper. This will turn out to be instructive in understanding the origin of the decoupling of isocurvature from curvature modes in the 2-field case. The action is still
\begin{eqnarray}
S&=&\int d^4 x\sqrt{-g}[\frac{1}{2}M^2({\vec \phi}) R-\frac{1}{2}\sum_{A=1}^2\nabla_\mu\phi^A\nabla^\mu \phi^A - W({\vec \phi})]
\nn \\
\label{SAp}
\end{eqnarray}
where we now have $M^2=-\frac{1}{6}\sum_{A=1}^3\alpha_A (\phi^A)^2$ and
$W({\vec \phi})   =  \sum_{A,B=1}^3\lambda_{AB}(\phi^A)^2(\phi^B)^2$. As before, a crucial quantity now is the turn rate $\omega^A$;  whenever $\omega^A = 0$, curvature and isocurvature modes decouple \cite{Kaiser:2012ak}. While we indeed found $\omega^A = 0$ in the 2-field case, the 3-field case is significantly different. One first new feature relevant to the computation of $\omega^A$ is that, while scale-invariance still enforces $X_A\phi^A = 0$, this no longer in general eliminates all $\lambda_{AB}$ dependence from $\hat X_A$. We therefore here, for simplicity, choose to set all parameters in the potential except for $\lambda_{22}$ to zero; a choice that will be sufficient to show that generically $\omega^A \neq 0$ in multi-field extensions of the 2-field model considered in the main text.  Explicitly calculating $\omega^A$ for the 3-field model in question, we then obtain
\begin{align}
\omega_1 &= \frac{\alpha_1 (1 - \alpha_3) \phi_1{}^2 + (1 - \alpha_2) \alpha_3 \phi_2{}^2 + (1 - \alpha_3) \alpha_3 \phi_3{}^2}{(\alpha_1 -  \alpha_3) \phi_1{} \phi_2{}{\cal A}}, \nn \\
\omega_2 &= \frac{1}{{\cal A}}, \nn \\
\omega_3 &= \frac{(\alpha_1-1) \alpha_1 \phi_1{}^2 + \alpha_1 (\alpha_2-1) \phi_2{}^2 + ( \alpha_1-1) \alpha_3 \phi_3{}^2}{(\alpha_1 -  \alpha_3) \phi_2{} \phi_3{}{\cal A}}, 
\end{align}
where we have written $\omega^A \equiv \{\omega_1, \omega_2,\omega_3\}$ and have defined the following shorthand notation
{
\begin{align}
{\cal A}^2 &\equiv - \frac{27 \bigl(\sum_{A=1}^3\alpha_A(\alpha_A-1)\phi_A^2\bigr)^3 {\cal B}^3}{2 \alpha_1^2 (\alpha_2-1)^4 (\alpha_1 -  \alpha_3)^4 \alpha_3^2 \phi_1{}^4 \phi_2{}^4 \phi_3{}^4 (-6M^2)^4}, \nn \\
{\cal B} &\equiv \alpha_1^3 \phi_1{}^4 + \alpha_1 (\alpha_3-2) \alpha_3 \phi_1{}^2 \phi_3{}^2 + \alpha_3^2 \phi_3{}^2 \bigl((\alpha_2-1) \phi_2{}^2 \nn \\ &+ (\alpha_3-1) \phi_3{}^2\bigr) -  \alpha_1^2 \phi_1{}^2 (\phi_1{}^2 + \phi_2{}^2 -  \alpha_2 \phi_2{}^2 -  \alpha_3 \phi_3{}^2),
\end{align}
}
using that $M^2=-\frac{1}{6}\sum_{A=1}^3\alpha_A (\phi^A)^2$ as before.
Given these expressions, we can then succintly express the magnitude of the turn rate $\omega^2 = |\omega_A \omega^A|$ as 
{
%\begin{align}
%\omega^2 = \frac{4 \alpha_1^2 (\alpha_2 - 1)^4 (\alpha_1 -  \alpha_3)^2 \alpha_3^2 \phi_1{}^2 \phi_2{}^2 \phi_3{}^2 {K} (-6M^2)^3}{9 \Bigl[\sum_{A=1}^3\alpha_A(\alpha_A-1)\phi_A^2\Bigr]^3 {\cal B}^2},
%\end{align}
\begin{align}
\omega^2 = \left|\frac{96 \alpha_1^2 (\alpha_2 - 1)^4 (\alpha_1 -  \alpha_3)^2 \alpha_3^2 \phi_1{}^2 \phi_2{}^2 \phi_3{}^2 {K} M^6}{\Bigl[\sum_{A=1}^3\alpha_A(\alpha_A-1)\phi_A^2\Bigr]^3 {\cal B}^2}\right|,
\end{align}
}
where $K$, in analogy to the constant from equation \ref{ellipse}, satisfies
\begin{eqnarray}
K=\frac{1}{2}\sum_{A=1}^3(1-\alpha_A)(\phi^A)^2 \label{hyperellipsoid}
\end{eqnarray}
and describes the hyper-ellipsoide constraint surface in which the inflationary trajectory is embedded.
Clearly we therefore have a non-zero turn rate and associated mixing between curvature and isocurvature modes. This shows that the decoupling of these modes from one another cannot be a general consequence of scale invariance, irrespective of field-space dimension. In the three-dimensional case, we now have an isocurvature mode, orthogonal to the scalar field trajectory, which lies on the constraint surface. Further work needs to be done to asses if this isocurvature mode is long-lived.

\newpage

%===============================================================================
% BIBLIOGRAPHY
%===============================================================================

\bibliographystyle{apsrev4-1}
\bibliography{InfSI}

\begin{thebibliography}{40}%
\makeatletter
\providecommand \@ifxundefined [1]{%
 \@ifx{#1\undefined}
}%
\providecommand \@ifnum [1]{%
 \ifnum #1\expandafter \@firstoftwo
 \else \expandafter \@secondoftwo
 \fi
}%
\providecommand \@ifx [1]{%
 \ifx #1\expandafter \@firstoftwo
 \else \expandafter \@secondoftwo
 \fi
}%
\providecommand \natexlab [1]{#1}%
\providecommand \enquote  [1]{``#1''}%
\providecommand \bibnamefont  [1]{#1}%
\providecommand \bibfnamefont [1]{#1}%
\providecommand \citenamefont [1]{#1}%
\providecommand \href@noop [0]{\@secondoftwo}%
\providecommand \href [0]{\begingroup \@sanitize@url \@href}%
\providecommand \@href[1]{\@@startlink{#1}\@@href}%
\providecommand \@@href[1]{\endgroup#1\@@endlink}%
\providecommand \@sanitize@url [0]{\catcode `\\12\catcode `\$12\catcode
  `\&12\catcode `\#12\catcode `\^12\catcode `\_12\catcode `\%12\relax}%
\providecommand \@@startlink[1]{}%
\providecommand \@@endlink[0]{}%
\providecommand \url  [0]{\begingroup\@sanitize@url \@url }%
\providecommand \@url [1]{\endgroup\@href {#1}{\urlprefix }}%
\providecommand \urlprefix  [0]{URL }%
\providecommand \Eprint [0]{\href }%
\providecommand \doibase [0]{http://dx.doi.org/}%
\providecommand \selectlanguage [0]{\@gobble}%
\providecommand \bibinfo  [0]{\@secondoftwo}%
\providecommand \bibfield  [0]{\@secondoftwo}%
\providecommand \translation [1]{[#1]}%
\providecommand \BibitemOpen [0]{}%
\providecommand \bibitemStop [0]{}%
\providecommand \bibitemNoStop [0]{.\EOS\space}%
\providecommand \EOS [0]{\spacefactor3000\relax}%
\providecommand \BibitemShut  [1]{\csname bibitem#1\endcsname}%
\let\auto@bib@innerbib\@empty
%</preamble>
\bibitem [{\citenamefont {Ferreira}\ \emph
  {et~al.}(2016{\natexlab{a}})\citenamefont {Ferreira}, \citenamefont {Hill},\
  and\ \citenamefont {Ross}}]{Ferreira:2016vsc}%
  \BibitemOpen
  \bibfield  {author} {\bibinfo {author} {\bibfnamefont {P.~G.}\ \bibnamefont
  {Ferreira}}, \bibinfo {author} {\bibfnamefont {C.~T.}\ \bibnamefont {Hill}},
  \ and\ \bibinfo {author} {\bibfnamefont {G.~G.}\ \bibnamefont {Ross}},\
  }\href {\doibase 10.1016/j.physletb.2016.10.036} {\bibfield  {journal}
  {\bibinfo  {journal} {Phys. Lett. B}\ } (\bibinfo {year}
  {2016}{\natexlab{a}}),\ 10.1016/j.physletb.2016.10.036},\ \Eprint
  {http://arxiv.org/abs/1603.05983} {arXiv:1603.05983 [hep-th]} \BibitemShut
  {NoStop}%
%%CITATION = ARXIV:1603.05983;%%
\bibitem [{\citenamefont {Ferreira}\ \emph
  {et~al.}(2016{\natexlab{b}})\citenamefont {Ferreira}, \citenamefont {Hill},\
  and\ \citenamefont {Ross}}]{Ferreira:2016wem}%
  \BibitemOpen
  \bibfield  {author} {\bibinfo {author} {\bibfnamefont {P.~G.}\ \bibnamefont
  {Ferreira}}, \bibinfo {author} {\bibfnamefont {C.~T.}\ \bibnamefont {Hill}},
  \ and\ \bibinfo {author} {\bibfnamefont {G.~G.}\ \bibnamefont {Ross}},\
  }\href@noop {} {\  (\bibinfo {year} {2016}{\natexlab{b}})},\ \Eprint
  {http://arxiv.org/abs/1610.09243} {arXiv:1610.09243 [hep-th]} \BibitemShut
  {NoStop}%
%%CITATION = ARXIV:1610.09243;%%
\bibitem [{\citenamefont {Garcia-Bellido}\ \emph {et~al.}(2011)\citenamefont
  {Garcia-Bellido}, \citenamefont {Rubio}, \citenamefont {Shaposhnikov},\ and\
  \citenamefont {Zenhausern}}]{GarciaBellido:2011de}%
  \BibitemOpen
  \bibfield  {author} {\bibinfo {author} {\bibfnamefont {J.}~\bibnamefont
  {Garcia-Bellido}}, \bibinfo {author} {\bibfnamefont {J.}~\bibnamefont
  {Rubio}}, \bibinfo {author} {\bibfnamefont {M.}~\bibnamefont {Shaposhnikov}},
  \ and\ \bibinfo {author} {\bibfnamefont {D.}~\bibnamefont {Zenhausern}},\
  }\href {\doibase 10.1103/PhysRevD.84.123504} {\bibfield  {journal} {\bibinfo
  {journal} {Phys. Rev.}\ }\textbf {\bibinfo {volume} {D84}},\ \bibinfo {pages}
  {123504} (\bibinfo {year} {2011})},\ \Eprint {http://arxiv.org/abs/1107.2163}
  {arXiv:1107.2163 [hep-ph]} \BibitemShut {NoStop}%
%%CITATION = ARXIV:1107.2163;%%
\bibitem [{\citenamefont {Rubio}\ and\ \citenamefont
  {Shaposhnikov}(2014)}]{Rubio:2014wta}%
  \BibitemOpen
  \bibfield  {author} {\bibinfo {author} {\bibfnamefont {J.}~\bibnamefont
  {Rubio}}\ and\ \bibinfo {author} {\bibfnamefont {M.}~\bibnamefont
  {Shaposhnikov}},\ }\href {\doibase 10.1103/PhysRevD.90.027307} {\bibfield
  {journal} {\bibinfo  {journal} {Phys. Rev.}\ }\textbf {\bibinfo {volume}
  {D90}},\ \bibinfo {pages} {027307} (\bibinfo {year} {2014})},\ \Eprint
  {http://arxiv.org/abs/1406.5182} {arXiv:1406.5182 [hep-ph]} \BibitemShut
  {NoStop}%
%%CITATION = ARXIV:1406.5182;%%
\bibitem [{\citenamefont {Bezrukov}\ \emph {et~al.}(2015)\citenamefont
  {Bezrukov}, \citenamefont {Rubio},\ and\ \citenamefont
  {Shaposhnikov}}]{Bezrukov:2014ipa}%
  \BibitemOpen
  \bibfield  {author} {\bibinfo {author} {\bibfnamefont {F.}~\bibnamefont
  {Bezrukov}}, \bibinfo {author} {\bibfnamefont {J.}~\bibnamefont {Rubio}}, \
  and\ \bibinfo {author} {\bibfnamefont {M.}~\bibnamefont {Shaposhnikov}},\
  }\href {\doibase 10.1103/PhysRevD.92.083512} {\bibfield  {journal} {\bibinfo
  {journal} {Phys. Rev.}\ }\textbf {\bibinfo {volume} {D92}},\ \bibinfo {pages}
  {083512} (\bibinfo {year} {2015})},\ \Eprint {http://arxiv.org/abs/1412.3811}
  {arXiv:1412.3811 [hep-ph]} \BibitemShut {NoStop}%
%%CITATION = ARXIV:1412.3811;%%
\bibitem [{\citenamefont {Karananas}\ and\ \citenamefont
  {Shaposhnikov}(2016)}]{Karananas:2016grc}%
  \BibitemOpen
  \bibfield  {author} {\bibinfo {author} {\bibfnamefont {G.~K.}\ \bibnamefont
  {Karananas}}\ and\ \bibinfo {author} {\bibfnamefont {M.}~\bibnamefont
  {Shaposhnikov}},\ }\href {\doibase 10.1103/PhysRevD.93.084052} {\bibfield
  {journal} {\bibinfo  {journal} {Phys. Rev.}\ }\textbf {\bibinfo {volume}
  {D93}},\ \bibinfo {pages} {084052} (\bibinfo {year} {2016})},\ \Eprint
  {http://arxiv.org/abs/1603.01274} {arXiv:1603.01274 [hep-th]} \BibitemShut
  {NoStop}%
%%CITATION = ARXIV:1603.01274;%%
\bibitem [{\citenamefont {Wetterich}(1988)}]{Wetterich:1987fm}%
  \BibitemOpen
  \bibfield  {author} {\bibinfo {author} {\bibfnamefont {C.}~\bibnamefont
  {Wetterich}},\ }\href {\doibase 10.1016/0550-3213(88)90193-9} {\bibfield
  {journal} {\bibinfo  {journal} {Nucl. Phys.}\ }\textbf {\bibinfo {volume}
  {B302}},\ \bibinfo {pages} {668} (\bibinfo {year} {1988})},\ \Eprint
  {http://arxiv.org/abs/1711.03844} {arXiv:1711.03844 [hep-th]} \BibitemShut
  {NoStop}%
%%CITATION = ARXIV:1711.03844;%%
\bibitem [{\citenamefont {Kallosh}\ and\ \citenamefont
  {Linde}(2013)}]{Kallosh:2013maa}%
  \BibitemOpen
  \bibfield  {author} {\bibinfo {author} {\bibfnamefont {R.}~\bibnamefont
  {Kallosh}}\ and\ \bibinfo {author} {\bibfnamefont {A.}~\bibnamefont
  {Linde}},\ }\href {\doibase 10.1088/1475-7516/2013/10/033} {\bibfield
  {journal} {\bibinfo  {journal} {JCAP}\ }\textbf {\bibinfo {volume} {1310}},\
  \bibinfo {pages} {033} (\bibinfo {year} {2013})},\ \Eprint
  {http://arxiv.org/abs/1307.7938} {arXiv:1307.7938 [hep-th]} \BibitemShut
  {NoStop}%
%%CITATION = ARXIV:1307.7938;%%
\bibitem [{\citenamefont {Carrasco}\ \emph {et~al.}(2015)\citenamefont
  {Carrasco}, \citenamefont {Kallosh},\ and\ \citenamefont
  {Linde}}]{Carrasco:2015rva}%
  \BibitemOpen
  \bibfield  {author} {\bibinfo {author} {\bibfnamefont {J.~J.~M.}\
  \bibnamefont {Carrasco}}, \bibinfo {author} {\bibfnamefont {R.}~\bibnamefont
  {Kallosh}}, \ and\ \bibinfo {author} {\bibfnamefont {A.}~\bibnamefont
  {Linde}},\ }\href {\doibase 10.1103/PhysRevD.92.063519} {\bibfield  {journal}
  {\bibinfo  {journal} {Phys. Rev.}\ }\textbf {\bibinfo {volume} {D92}},\
  \bibinfo {pages} {063519} (\bibinfo {year} {2015})},\ \Eprint
  {http://arxiv.org/abs/1506.00936} {arXiv:1506.00936 [hep-th]} \BibitemShut
  {NoStop}%
%%CITATION = ARXIV:1506.00936;%%
\bibitem [{\citenamefont {Quiros}(2014)}]{Quiros:2014wda}%
  \BibitemOpen
  \bibfield  {author} {\bibinfo {author} {\bibfnamefont {I.}~\bibnamefont
  {Quiros}},\ }\href@noop {} {\  (\bibinfo {year} {2014})},\ \Eprint
  {http://arxiv.org/abs/1405.6668} {arXiv:1405.6668 [gr-qc]} \BibitemShut
  {NoStop}%
%%CITATION = ARXIV:1405.6668;%%
\bibitem [{\citenamefont {Kurkov}(2016)}]{Kurkov:2016zpd}%
  \BibitemOpen
  \bibfield  {author} {\bibinfo {author} {\bibfnamefont {M.}~\bibnamefont
  {Kurkov}},\ }\href {\doibase 10.1140/epjc/s10052-016-4178-6} {\bibfield
  {journal} {\bibinfo  {journal} {Eur. Phys. J.}\ }\textbf {\bibinfo {volume}
  {C76}},\ \bibinfo {pages} {329} (\bibinfo {year} {2016})},\ \Eprint
  {http://arxiv.org/abs/1601.00622} {arXiv:1601.00622 [hep-th]} \BibitemShut
  {NoStop}%
%%CITATION = ARXIV:1601.00622;%%
\bibitem [{\citenamefont {Karananas}\ and\ \citenamefont
  {Shaposhnikov}(2018)}]{Karananas:2017zrg}%
  \BibitemOpen
  \bibfield  {author} {\bibinfo {author} {\bibfnamefont {G.~K.}\ \bibnamefont
  {Karananas}}\ and\ \bibinfo {author} {\bibfnamefont {M.}~\bibnamefont
  {Shaposhnikov}},\ }\href {\doibase 10.1103/PhysRevD.97.045009} {\bibfield
  {journal} {\bibinfo  {journal} {Phys. Rev.}\ }\textbf {\bibinfo {volume}
  {D97}},\ \bibinfo {pages} {045009} (\bibinfo {year} {2018})},\ \Eprint
  {http://arxiv.org/abs/1708.02220} {arXiv:1708.02220 [hep-th]} \BibitemShut
  {NoStop}%
%%CITATION = ARXIV:1708.02220;%%
\bibitem [{\citenamefont {Karam}\ \emph {et~al.}(2017)\citenamefont {Karam},
  \citenamefont {Pappas},\ and\ \citenamefont {Tamvakis}}]{Karam:2017zno}%
  \BibitemOpen
  \bibfield  {author} {\bibinfo {author} {\bibfnamefont {A.}~\bibnamefont
  {Karam}}, \bibinfo {author} {\bibfnamefont {T.}~\bibnamefont {Pappas}}, \
  and\ \bibinfo {author} {\bibfnamefont {K.}~\bibnamefont {Tamvakis}},\ }\href
  {\doibase 10.1103/PhysRevD.96.064036} {\bibfield  {journal} {\bibinfo
  {journal} {Phys. Rev.}\ }\textbf {\bibinfo {volume} {D96}},\ \bibinfo {pages}
  {064036} (\bibinfo {year} {2017})},\ \Eprint
  {http://arxiv.org/abs/1707.00984} {arXiv:1707.00984 [gr-qc]} \BibitemShut
  {NoStop}%
%%CITATION = ARXIV:1707.00984;%%
\bibitem [{\citenamefont {Rubio}\ and\ \citenamefont
  {Wetterich}(2017)}]{Rubio:2017gty}%
  \BibitemOpen
  \bibfield  {author} {\bibinfo {author} {\bibfnamefont {J.}~\bibnamefont
  {Rubio}}\ and\ \bibinfo {author} {\bibfnamefont {C.}~\bibnamefont
  {Wetterich}},\ }\href {\doibase 10.1103/PhysRevD.96.063509} {\bibfield
  {journal} {\bibinfo  {journal} {Phys. Rev.}\ }\textbf {\bibinfo {volume}
  {D96}},\ \bibinfo {pages} {063509} (\bibinfo {year} {2017})},\ \Eprint
  {http://arxiv.org/abs/1705.00552} {arXiv:1705.00552 [gr-qc]} \BibitemShut
  {NoStop}%
%%CITATION = ARXIV:1705.00552;%%
\bibitem [{\citenamefont {Karananas}\ and\ \citenamefont
  {Shaposhnikov}(2017)}]{Karananas:2017mxm}%
  \BibitemOpen
  \bibfield  {author} {\bibinfo {author} {\bibfnamefont {G.~K.}\ \bibnamefont
  {Karananas}}\ and\ \bibinfo {author} {\bibfnamefont {M.}~\bibnamefont
  {Shaposhnikov}},\ }\href {\doibase 10.1016/j.physletb.2017.05.065} {\bibfield
   {journal} {\bibinfo  {journal} {Phys. Lett.}\ }\textbf {\bibinfo {volume}
  {B771}},\ \bibinfo {pages} {332} (\bibinfo {year} {2017})},\ \Eprint
  {http://arxiv.org/abs/1703.02964} {arXiv:1703.02964 [hep-ph]} \BibitemShut
  {NoStop}%
%%CITATION = ARXIV:1703.02964;%%
\bibitem [{\citenamefont {Kannike}\ \emph {et~al.}(2017)\citenamefont
  {Kannike}, \citenamefont {Raidal}, \citenamefont {Spethmann},\ and\
  \citenamefont {Veermäe}}]{Kannike:2016wuy}%
  \BibitemOpen
  \bibfield  {author} {\bibinfo {author} {\bibfnamefont {K.}~\bibnamefont
  {Kannike}}, \bibinfo {author} {\bibfnamefont {M.}~\bibnamefont {Raidal}},
  \bibinfo {author} {\bibfnamefont {C.}~\bibnamefont {Spethmann}}, \ and\
  \bibinfo {author} {\bibfnamefont {H.}~\bibnamefont {Veermäe}},\ }\href
  {\doibase 10.1007/JHEP04(2017)026} {\bibfield  {journal} {\bibinfo  {journal}
  {JHEP}\ }\textbf {\bibinfo {volume} {04}},\ \bibinfo {pages} {026} (\bibinfo
  {year} {2017})},\ \Eprint {http://arxiv.org/abs/1610.06571} {arXiv:1610.06571
  [hep-ph]} \BibitemShut {NoStop}%
%%CITATION = ARXIV:1610.06571;%%
\bibitem [{\citenamefont {Einhorn}\ and\ \citenamefont
  {Jones}(2017)}]{Einhorn:2017icw}%
  \BibitemOpen
  \bibfield  {author} {\bibinfo {author} {\bibfnamefont {M.~B.}\ \bibnamefont
  {Einhorn}}\ and\ \bibinfo {author} {\bibfnamefont {D.~R.~T.}\ \bibnamefont
  {Jones}},\ }\href {\doibase 10.1103/PhysRevD.96.124025} {\bibfield  {journal}
  {\bibinfo  {journal} {Phys. Rev.}\ }\textbf {\bibinfo {volume} {D96}},\
  \bibinfo {pages} {124025} (\bibinfo {year} {2017})},\ \Eprint
  {http://arxiv.org/abs/1710.03795} {arXiv:1710.03795 [hep-th]} \BibitemShut
  {NoStop}%
%%CITATION = ARXIV:1710.03795;%%
\bibitem [{\citenamefont {Salvio}\ and\ \citenamefont
  {Strumia}(2018)}]{Salvio:2017qkx}%
  \BibitemOpen
  \bibfield  {author} {\bibinfo {author} {\bibfnamefont {A.}~\bibnamefont
  {Salvio}}\ and\ \bibinfo {author} {\bibfnamefont {A.}~\bibnamefont
  {Strumia}},\ }\href {\doibase 10.1140/epjc/s10052-018-5588-4} {\bibfield
  {journal} {\bibinfo  {journal} {Eur. Phys. J.}\ }\textbf {\bibinfo {volume}
  {C78}},\ \bibinfo {pages} {124} (\bibinfo {year} {2018})},\ \Eprint
  {http://arxiv.org/abs/1705.03896} {arXiv:1705.03896 [hep-th]} \BibitemShut
  {NoStop}%
%%CITATION = ARXIV:1705.03896;%%
\bibitem [{\citenamefont {Salvio}(2017)}]{Salvio:2017xul}%
  \BibitemOpen
  \bibfield  {author} {\bibinfo {author} {\bibfnamefont {A.}~\bibnamefont
  {Salvio}},\ }\href {\doibase 10.1140/epjc/s10052-017-4825-6} {\bibfield
  {journal} {\bibinfo  {journal} {Eur. Phys. J.}\ }\textbf {\bibinfo {volume}
  {C77}},\ \bibinfo {pages} {267} (\bibinfo {year} {2017})},\ \Eprint
  {http://arxiv.org/abs/1703.08012} {arXiv:1703.08012 [astro-ph.CO]}
  \BibitemShut {NoStop}%
%%CITATION = ARXIV:1703.08012;%%
\bibitem [{\citenamefont {Salvio}\ and\ \citenamefont
  {Strumia}(2014)}]{Salvio:2014soa}%
  \BibitemOpen
  \bibfield  {author} {\bibinfo {author} {\bibfnamefont {A.}~\bibnamefont
  {Salvio}}\ and\ \bibinfo {author} {\bibfnamefont {A.}~\bibnamefont
  {Strumia}},\ }\href {\doibase 10.1007/JHEP06(2014)080} {\bibfield  {journal}
  {\bibinfo  {journal} {JHEP}\ }\textbf {\bibinfo {volume} {06}},\ \bibinfo
  {pages} {080} (\bibinfo {year} {2014})},\ \Eprint
  {http://arxiv.org/abs/1403.4226} {arXiv:1403.4226 [hep-ph]} \BibitemShut
  {NoStop}%
%%CITATION = ARXIV:1403.4226;%%
\bibitem [{\citenamefont {Gordon}\ \emph {et~al.}(2001)\citenamefont {Gordon},
  \citenamefont {Wands}, \citenamefont {Bassett},\ and\ \citenamefont
  {Maartens}}]{Gordon:2000hv}%
  \BibitemOpen
  \bibfield  {author} {\bibinfo {author} {\bibfnamefont {C.}~\bibnamefont
  {Gordon}}, \bibinfo {author} {\bibfnamefont {D.}~\bibnamefont {Wands}},
  \bibinfo {author} {\bibfnamefont {B.~A.}\ \bibnamefont {Bassett}}, \ and\
  \bibinfo {author} {\bibfnamefont {R.}~\bibnamefont {Maartens}},\ }\href
  {\doibase 10.1103/PhysRevD.63.023506} {\bibfield  {journal} {\bibinfo
  {journal} {Phys. Rev.}\ }\textbf {\bibinfo {volume} {D63}},\ \bibinfo {pages}
  {023506} (\bibinfo {year} {2001})},\ \Eprint
  {http://arxiv.org/abs/astro-ph/0009131} {arXiv:astro-ph/0009131 [astro-ph]}
  \BibitemShut {NoStop}%
%%CITATION = ASTRO-PH/0009131;%%
\bibitem [{\citenamefont {Bartolo}\ \emph {et~al.}(2004)\citenamefont
  {Bartolo}, \citenamefont {Komatsu}, \citenamefont {Matarrese},\ and\
  \citenamefont {Riotto}}]{Bartolo:2004if}%
  \BibitemOpen
  \bibfield  {author} {\bibinfo {author} {\bibfnamefont {N.}~\bibnamefont
  {Bartolo}}, \bibinfo {author} {\bibfnamefont {E.}~\bibnamefont {Komatsu}},
  \bibinfo {author} {\bibfnamefont {S.}~\bibnamefont {Matarrese}}, \ and\
  \bibinfo {author} {\bibfnamefont {A.}~\bibnamefont {Riotto}},\ }\href
  {\doibase 10.1016/j.physrep.2004.08.022} {\bibfield  {journal} {\bibinfo
  {journal} {Phys. Rept.}\ }\textbf {\bibinfo {volume} {402}},\ \bibinfo
  {pages} {103} (\bibinfo {year} {2004})},\ \Eprint
  {http://arxiv.org/abs/astro-ph/0406398} {arXiv:astro-ph/0406398 [astro-ph]}
  \BibitemShut {NoStop}%
%%CITATION = ASTRO-PH/0406398;%%
\bibitem [{\citenamefont {Karamitsos}\ and\ \citenamefont
  {Pilaftsis}(2018)}]{KP}%
  \BibitemOpen
  \bibfield  {author} {\bibinfo {author} {\bibfnamefont {S.}~\bibnamefont
  {Karamitsos}}\ and\ \bibinfo {author} {\bibfnamefont {A.}~\bibnamefont
  {Pilaftsis}},\ }\href {\doibase 10.1016/j.nuclphysb.2017.12.015} {\bibfield
  {journal} {\bibinfo  {journal} {Nucl. Phys.}\ }\textbf {\bibinfo {volume}
  {B927}},\ \bibinfo {pages} {219} (\bibinfo {year} {2018})},\ \Eprint
  {http://arxiv.org/abs/1706.07011} {arXiv:1706.07011 [hep-ph]} \BibitemShut
  {NoStop}%
%%CITATION = ARXIV:1706.07011;%%
\bibitem [{\citenamefont {Ferreira}\ \emph {et~al.}(2018)\citenamefont
  {Ferreira}, \citenamefont {Hill},\ and\ \citenamefont
  {Ross}}]{Ferreira:2018itt}%
  \BibitemOpen
  \bibfield  {author} {\bibinfo {author} {\bibfnamefont {P.~G.}\ \bibnamefont
  {Ferreira}}, \bibinfo {author} {\bibfnamefont {C.~T.}\ \bibnamefont {Hill}},
  \ and\ \bibinfo {author} {\bibfnamefont {G.~G.}\ \bibnamefont {Ross}},\
  }\href@noop {} {\  (\bibinfo {year} {2018})},\ \Eprint
  {http://arxiv.org/abs/1801.07676} {arXiv:1801.07676 [hep-th]} \BibitemShut
  {NoStop}%
%%CITATION = ARXIV:1801.07676;%%
\bibitem [{\citenamefont {Ferreira}\ \emph {et~al.}(2017)\citenamefont
  {Ferreira}, \citenamefont {Hill},\ and\ \citenamefont
  {Ross}}]{Ferreira:2016kxi}%
  \BibitemOpen
  \bibfield  {author} {\bibinfo {author} {\bibfnamefont {P.~G.}\ \bibnamefont
  {Ferreira}}, \bibinfo {author} {\bibfnamefont {C.~T.}\ \bibnamefont {Hill}},
  \ and\ \bibinfo {author} {\bibfnamefont {G.~G.}\ \bibnamefont {Ross}},\
  }\href {\doibase 10.1103/PhysRevD.95.064038} {\bibfield  {journal} {\bibinfo
  {journal} {Phys. Rev.}\ }\textbf {\bibinfo {volume} {D95}},\ \bibinfo {pages}
  {064038} (\bibinfo {year} {2017})},\ \Eprint
  {http://arxiv.org/abs/1612.03157} {arXiv:1612.03157 [gr-qc]} \BibitemShut
  {NoStop}%
%%CITATION = ARXIV:1612.03157;%%
\bibitem [{\citenamefont {Sasaki}\ and\ \citenamefont
  {Stewart}(1996)}]{Sasaki:1995aw}%
  \BibitemOpen
  \bibfield  {author} {\bibinfo {author} {\bibfnamefont {M.}~\bibnamefont
  {Sasaki}}\ and\ \bibinfo {author} {\bibfnamefont {E.~D.}\ \bibnamefont
  {Stewart}},\ }\href {\doibase 10.1143/PTP.95.71} {\bibfield  {journal}
  {\bibinfo  {journal} {Prog. Theor. Phys.}\ }\textbf {\bibinfo {volume}
  {95}},\ \bibinfo {pages} {71} (\bibinfo {year} {1996})},\ \Eprint
  {http://arxiv.org/abs/astro-ph/9507001} {arXiv:astro-ph/9507001 [astro-ph]}
  \BibitemShut {NoStop}%
%%CITATION = ASTRO-PH/9507001;%%
\bibitem [{\citenamefont {Elliston}\ \emph {et~al.}(2011)\citenamefont
  {Elliston}, \citenamefont {Mulryne}, \citenamefont {Seery},\ and\
  \citenamefont {Tavakol}}]{Elliston:2011et}%
  \BibitemOpen
  \bibfield  {author} {\bibinfo {author} {\bibfnamefont {J.}~\bibnamefont
  {Elliston}}, \bibinfo {author} {\bibfnamefont {D.}~\bibnamefont {Mulryne}},
  \bibinfo {author} {\bibfnamefont {D.}~\bibnamefont {Seery}}, \ and\ \bibinfo
  {author} {\bibfnamefont {R.}~\bibnamefont {Tavakol}},\ }\href {\doibase
  10.1142/S0217751X11054280, 10.1142/S2010194511001292} {\bibfield  {journal}
  {\bibinfo  {journal} {Int. J. Mod. Phys.}\ }\textbf {\bibinfo {volume}
  {A26}},\ \bibinfo {pages} {3821} (\bibinfo {year} {2011})},\ \bibinfo {note}
  {[Int. J. Mod. Phys. Conf. Ser.03,203(2011)]},\ \Eprint
  {http://arxiv.org/abs/1107.2270} {arXiv:1107.2270 [astro-ph.CO]} \BibitemShut
  {NoStop}%
%%CITATION = ARXIV:1107.2270;%%
\bibitem [{\citenamefont {Kaiser}\ and\ \citenamefont
  {Sfakianakis}(2014)}]{Kaiser:2013sna}%
  \BibitemOpen
  \bibfield  {author} {\bibinfo {author} {\bibfnamefont {D.~I.}\ \bibnamefont
  {Kaiser}}\ and\ \bibinfo {author} {\bibfnamefont {E.~I.}\ \bibnamefont
  {Sfakianakis}},\ }\href {\doibase 10.1103/PhysRevLett.112.011302} {\bibfield
  {journal} {\bibinfo  {journal} {Phys. Rev. Lett.}\ }\textbf {\bibinfo
  {volume} {112}},\ \bibinfo {pages} {011302} (\bibinfo {year} {2014})},\
  \Eprint {http://arxiv.org/abs/1304.0363} {arXiv:1304.0363 [astro-ph.CO]}
  \BibitemShut {NoStop}%
%%CITATION = ARXIV:1304.0363;%%
\bibitem [{\citenamefont {Achucarro}\ \emph {et~al.}(2013)\citenamefont
  {Achucarro}, \citenamefont {Gong}, \citenamefont {Palma},\ and\ \citenamefont
  {Patil}}]{Achucarro:2012fd}%
  \BibitemOpen
  \bibfield  {author} {\bibinfo {author} {\bibfnamefont {A.}~\bibnamefont
  {Achucarro}}, \bibinfo {author} {\bibfnamefont {J.-O.}\ \bibnamefont {Gong}},
  \bibinfo {author} {\bibfnamefont {G.~A.}\ \bibnamefont {Palma}}, \ and\
  \bibinfo {author} {\bibfnamefont {S.~P.}\ \bibnamefont {Patil}},\ }\href
  {\doibase 10.1103/PhysRevD.87.121301} {\bibfield  {journal} {\bibinfo
  {journal} {Phys. Rev.}\ }\textbf {\bibinfo {volume} {D87}},\ \bibinfo {pages}
  {121301} (\bibinfo {year} {2013})},\ \Eprint {http://arxiv.org/abs/1211.5619}
  {arXiv:1211.5619 [astro-ph.CO]} \BibitemShut {NoStop}%
%%CITATION = ARXIV:1211.5619;%%
\bibitem [{\citenamefont {Gong}\ and\ \citenamefont
  {Tanaka}(2011)}]{Gong:2011uw}%
  \BibitemOpen
  \bibfield  {author} {\bibinfo {author} {\bibfnamefont {J.-O.}\ \bibnamefont
  {Gong}}\ and\ \bibinfo {author} {\bibfnamefont {T.}~\bibnamefont {Tanaka}},\
  }\href {\doibase 10.1088/1475-7516/2012/02/E01,
  10.1088/1475-7516/2011/03/015} {\bibfield  {journal} {\bibinfo  {journal}
  {JCAP}\ }\textbf {\bibinfo {volume} {1103}},\ \bibinfo {pages} {015}
  (\bibinfo {year} {2011})},\ \bibinfo {note} {[Erratum: JCAP1202,E01(2012)]},\
  \Eprint {http://arxiv.org/abs/1101.4809} {arXiv:1101.4809 [astro-ph.CO]}
  \BibitemShut {NoStop}%
%%CITATION = ARXIV:1101.4809;%%
\bibitem [{\citenamefont {White}\ \emph {et~al.}(2012)\citenamefont {White},
  \citenamefont {Minamitsuji},\ and\ \citenamefont {Sasaki}}]{White:2012ya}%
  \BibitemOpen
  \bibfield  {author} {\bibinfo {author} {\bibfnamefont {J.}~\bibnamefont
  {White}}, \bibinfo {author} {\bibfnamefont {M.}~\bibnamefont {Minamitsuji}},
  \ and\ \bibinfo {author} {\bibfnamefont {M.}~\bibnamefont {Sasaki}},\ }\href
  {\doibase 10.1088/1475-7516/2012/07/039} {\bibfield  {journal} {\bibinfo
  {journal} {JCAP}\ }\textbf {\bibinfo {volume} {1207}},\ \bibinfo {pages}
  {039} (\bibinfo {year} {2012})},\ \Eprint {http://arxiv.org/abs/1205.0656}
  {arXiv:1205.0656 [astro-ph.CO]} \BibitemShut {NoStop}%
%%CITATION = ARXIV:1205.0656;%%
\bibitem [{\citenamefont {Kaiser}\ \emph {et~al.}(2013)\citenamefont {Kaiser},
  \citenamefont {Mazenc},\ and\ \citenamefont {Sfakianakis}}]{Kaiser:2012ak}%
  \BibitemOpen
  \bibfield  {author} {\bibinfo {author} {\bibfnamefont {D.~I.}\ \bibnamefont
  {Kaiser}}, \bibinfo {author} {\bibfnamefont {E.~A.}\ \bibnamefont {Mazenc}},
  \ and\ \bibinfo {author} {\bibfnamefont {E.~I.}\ \bibnamefont
  {Sfakianakis}},\ }\href {\doibase 10.1103/PhysRevD.87.064004} {\bibfield
  {journal} {\bibinfo  {journal} {Phys. Rev.}\ }\textbf {\bibinfo {volume}
  {D87}},\ \bibinfo {pages} {064004} (\bibinfo {year} {2013})},\ \Eprint
  {http://arxiv.org/abs/1210.7487} {arXiv:1210.7487 [astro-ph.CO]} \BibitemShut
  {NoStop}%
%%CITATION = ARXIV:1210.7487;%%
\bibitem [{\citenamefont {Ade}\ \emph {et~al.}(2016)\citenamefont {Ade} \emph
  {et~al.}}]{PlanckCosmo2015}%
  \BibitemOpen
  \bibfield  {author} {\bibinfo {author} {\bibfnamefont {P.~A.~R.}\
  \bibnamefont {Ade}} \emph {et~al.} (\bibinfo {collaboration} {Planck}),\
  }\href {\doibase 10.1051/0004-6361/201525830} {\bibfield  {journal} {\bibinfo
   {journal} {Astron. Astrophys.}\ }\textbf {\bibinfo {volume} {594}},\
  \bibinfo {pages} {A13} (\bibinfo {year} {2016})},\ \Eprint
  {http://arxiv.org/abs/1502.01589} {arXiv:1502.01589 [astro-ph.CO]}
  \BibitemShut {NoStop}%
%%CITATION = ARXIV:1502.01589;%%
\bibitem [{\citenamefont {Maldacena}(2003)}]{Maldacena:2002vr}%
  \BibitemOpen
  \bibfield  {author} {\bibinfo {author} {\bibfnamefont {J.~M.}\ \bibnamefont
  {Maldacena}},\ }\href {\doibase 10.1088/1126-6708/2003/05/013} {\bibfield
  {journal} {\bibinfo  {journal} {JHEP}\ }\textbf {\bibinfo {volume} {05}},\
  \bibinfo {pages} {013} (\bibinfo {year} {2003})},\ \Eprint
  {http://arxiv.org/abs/astro-ph/0210603} {arXiv:astro-ph/0210603 [astro-ph]}
  \BibitemShut {NoStop}%
%%CITATION = ASTRO-PH/0210603;%%
\bibitem [{\citenamefont {Creminelli}\ and\ \citenamefont
  {Zaldarriaga}(2004)}]{Creminelli:2004yq}%
  \BibitemOpen
  \bibfield  {author} {\bibinfo {author} {\bibfnamefont {P.}~\bibnamefont
  {Creminelli}}\ and\ \bibinfo {author} {\bibfnamefont {M.}~\bibnamefont
  {Zaldarriaga}},\ }\href {\doibase 10.1088/1475-7516/2004/10/006} {\bibfield
  {journal} {\bibinfo  {journal} {JCAP}\ }\textbf {\bibinfo {volume} {0410}},\
  \bibinfo {pages} {006} (\bibinfo {year} {2004})},\ \Eprint
  {http://arxiv.org/abs/astro-ph/0407059} {arXiv:astro-ph/0407059 [astro-ph]}
  \BibitemShut {NoStop}%
%%CITATION = ASTRO-PH/0407059;%%
\bibitem [{\citenamefont {Elliston}\ \emph {et~al.}(2012)\citenamefont
  {Elliston}, \citenamefont {Seery},\ and\ \citenamefont
  {Tavakol}}]{Elliston:2012ab}%
  \BibitemOpen
  \bibfield  {author} {\bibinfo {author} {\bibfnamefont {J.}~\bibnamefont
  {Elliston}}, \bibinfo {author} {\bibfnamefont {D.}~\bibnamefont {Seery}}, \
  and\ \bibinfo {author} {\bibfnamefont {R.}~\bibnamefont {Tavakol}},\ }\href
  {\doibase 10.1088/1475-7516/2012/11/060} {\bibfield  {journal} {\bibinfo
  {journal} {JCAP}\ }\textbf {\bibinfo {volume} {1211}},\ \bibinfo {pages}
  {060} (\bibinfo {year} {2012})},\ \Eprint {http://arxiv.org/abs/1208.6011}
  {arXiv:1208.6011 [astro-ph.CO]} \BibitemShut {NoStop}%
%%CITATION = ARXIV:1208.6011;%%
\bibitem [{\citenamefont {Byrnes}\ and\ \citenamefont
  {Gong}(2013)}]{Byrnes:2012sc}%
  \BibitemOpen
  \bibfield  {author} {\bibinfo {author} {\bibfnamefont {C.~T.}\ \bibnamefont
  {Byrnes}}\ and\ \bibinfo {author} {\bibfnamefont {J.-O.}\ \bibnamefont
  {Gong}},\ }\href {\doibase 10.1016/j.physletb.2012.11.052} {\bibfield
  {journal} {\bibinfo  {journal} {Phys. Lett.}\ }\textbf {\bibinfo {volume}
  {B718}},\ \bibinfo {pages} {718} (\bibinfo {year} {2013})},\ \Eprint
  {http://arxiv.org/abs/1210.1851} {arXiv:1210.1851 [astro-ph.CO]} \BibitemShut
  {NoStop}%
%%CITATION = ARXIV:1210.1851;%%
\bibitem [{\citenamefont {Lyth}\ and\ \citenamefont
  {Rodriguez}(2005)}]{Lyth:2005fi}%
  \BibitemOpen
  \bibfield  {author} {\bibinfo {author} {\bibfnamefont {D.~H.}\ \bibnamefont
  {Lyth}}\ and\ \bibinfo {author} {\bibfnamefont {Y.}~\bibnamefont
  {Rodriguez}},\ }\href {\doibase 10.1103/PhysRevLett.95.121302} {\bibfield
  {journal} {\bibinfo  {journal} {Phys. Rev. Lett.}\ }\textbf {\bibinfo
  {volume} {95}},\ \bibinfo {pages} {121302} (\bibinfo {year} {2005})},\
  \Eprint {http://arxiv.org/abs/astro-ph/0504045} {arXiv:astro-ph/0504045
  [astro-ph]} \BibitemShut {NoStop}%
%%CITATION = ASTRO-PH/0504045;%%
\bibitem [{\citenamefont {Kenton}\ and\ \citenamefont
  {Mulryne}(2015)}]{Kenton:2015lxa}%
  \BibitemOpen
  \bibfield  {author} {\bibinfo {author} {\bibfnamefont {Z.}~\bibnamefont
  {Kenton}}\ and\ \bibinfo {author} {\bibfnamefont {D.~J.}\ \bibnamefont
  {Mulryne}},\ }\href {\doibase 10.1088/1475-7516/2015/10/018} {\bibfield
  {journal} {\bibinfo  {journal} {JCAP}\ }\textbf {\bibinfo {volume} {1510}},\
  \bibinfo {pages} {018} (\bibinfo {year} {2015})},\ \Eprint
  {http://arxiv.org/abs/1507.08629} {arXiv:1507.08629 [astro-ph.CO]}
  \BibitemShut {NoStop}%
%%CITATION = ARXIV:1507.08629;%%
\bibitem [{\citenamefont {Acquaviva}\ \emph {et~al.}(2003)\citenamefont
  {Acquaviva}, \citenamefont {Bartolo}, \citenamefont {Matarrese},\ and\
  \citenamefont {Riotto}}]{Acquaviva:2002ud}%
  \BibitemOpen
  \bibfield  {author} {\bibinfo {author} {\bibfnamefont {V.}~\bibnamefont
  {Acquaviva}}, \bibinfo {author} {\bibfnamefont {N.}~\bibnamefont {Bartolo}},
  \bibinfo {author} {\bibfnamefont {S.}~\bibnamefont {Matarrese}}, \ and\
  \bibinfo {author} {\bibfnamefont {A.}~\bibnamefont {Riotto}},\ }\href
  {\doibase 10.1016/S0550-3213(03)00550-9} {\bibfield  {journal} {\bibinfo
  {journal} {Nucl. Phys.}\ }\textbf {\bibinfo {volume} {B667}},\ \bibinfo
  {pages} {119} (\bibinfo {year} {2003})},\ \Eprint
  {http://arxiv.org/abs/astro-ph/0209156} {arXiv:astro-ph/0209156 [astro-ph]}
  \BibitemShut {NoStop}%
%%CITATION = ASTRO-PH/0209156;%%
\end{thebibliography}%

\end{document}